\documentclass{elsart}
\usepackage{graphics,natbib,graphicx,psfig,amssymb} 
\journal{New Astronomy}
\input psfig.sty
\begin{document}
\begin{frontmatter}
\title{Spatial distribution and galactic model parameters of cataclysmic variables}
\author[istanbul]{T. Ak\corauthref{cor}},
\corauth[cor]{corresponding author.}
\ead{tanselak@istanbul.edu.tr}
\author[istanbul]{S. Bilir},
\author[istanbul]{S. Ak},
\author[canakkale,tug]{Z. Eker}
\address[istanbul]{Istanbul University, Faculty of Sciences, Department 
of Astronomy and Space Sciences, 34119 University, Istanbul, Turkey}
\address[canakkale]{\c Canakkale Onsekiz Mart University, Faculty of Sciences and 
Arts, Ulup\i nar Astrophysical Observatory, 17100, \c Canakkale, Turkey}
\address[tug]{T\"UB\.ITAK National Observatory, Akdeniz University Campus, 07058 Antalya, Turkey}

\begin{abstract}

The spatial distribution, galactic model parameters and luminosity function 
of cataclysmic variables (CVs) in the solar neighbourhood have been 
determined from a carefully established sample of 459 CVs. The sample contains 
all of the CVs with distances computed from the Period-Luminosity-Colours (PLCs)
relation of CVs which has been recently derived and calibrated with {\em 2MASS}
photometric data. It has been found that an exponential function fits best to the 
observational $z$-distributions of all of the CVs in the sample, non-magnetic 
CVs and dwarf novae, while the $sech^{2}$ function is more appropriate for 
nova-like stars and polars. The vertical scaleheight of CVs is 158$\pm$14 pc 
for the {\em 2MASS} $J$-band limiting apparent magnitude of 15.8. 
On the other hand, the vertical scaleheights are 128$\pm$20 and 
160$\pm$5 pc for dwarf novae and nova-like stars, respectively. The local space density 
of CVs is found to be $\sim3\times10^{-5}$ $pc^{-3}$ which is in agreement with the lower 
limit of the theoretical predictions. The luminosity function of CVs shows an increasing trend 
toward higher space densities at low luminosities, implying that the number of 
short-period systems should be high. 
The discrepancies between the theoretical and observational population studies of 
CVs will almost disappear if for the $z$-dependence of the space density the $sech^{2}$ 
density function is used. \\

\end{abstract}

\end{frontmatter}

\section{Introduction}

Cataclysmic variables (hereafter CVs) are short period semidetached interacting binary            
systems containing a white dwarf as a primary which accretes matter from a Roche-lobe-filled 
red dwarf companion \citep{Knigge2006} typically via an accretion disc. A bright spot 
is formed in the location where the matter stream impacts on the accretion disc. 
In the CVs with strong magnetic primaries, the magnetic field is strong enough 
to prevent the formation of an accretion disc. Disc formation could be prevented, 
but accretion continues through channels as accretion flows. For a detailed description 
of the CV phenomenon and its subclasses see \citet{Warner1995} and \citet{Hellier2001}.

Many basic properties of CVs, such as the masses of the component stars, cannot be 
directly determined \citep{SmithandDhillon1998}. The orbital period $P_{orb}$ is one 
of the exceptions as it is known directly from observations with a fair accuracy. 
Orbital periods with sufficient accuracy are available now for more than 640 
systems \citep{Downesetal2001,RitterandKolb2003}. Since an orbital period distribution 
is a useful indicator of dynamical evolution of orbits, most studies which are interested 
in the evolution of CVs are mainly concentrated on the distribution of orbital periods. The 
most striking features of the distribution of CV periods are the period gap and 
a sharp cut-off at about 80 min 
\citep{SpruitandRitter1983,Hameuryetal1988,HowellNelsonRappaport2001,Willemsetal2005,PretoriusKniggeKolb2007}.

Galactic space and velocity distributions could be crucial for various types of objects 
to indicate a general scenario or a stage on a scenario about their evolution. Any proposed 
evolutionary scheme must not be in conflict with data from stellar statistics \citep{Duerbeck1984}. 
If the space densities of systems in various intervals of orbital periods were known this would 
seriously improve our understanding of CV evolution.
However, observational selection effects are usually strong. Therefore, even a rough 
estimate of space density may help to constrain the population models \citep{Patterson1984}. 
Standard models for the population of CVs predict a galactic space density of $10^{-5}-10^{-4}$ pc$^{-3}$
\citep{RitterandBurkert1986,deKool1992,Politano1996,Kolb2001,Willemsetal2005,Willemsetal2006}.
These predictions are inconsistent with the observed space densities of $\sim$$10^{-6}$$-$$10^{-5}$ pc$^{-3}$
\citep{Warner1974,Patterson1984,Patterson1998,Ringwald1993,Schwopeetal2002,Aungwerojwitetal2005,
AraujoBetancoretal2005,Grindlayetal2005}.

From the observational point of view, the first step of deriving the space density of CVs is to 
collect the most truthfull CV distances in a selected sample. Only after reliable distances are available,
the observed surface angular density distribution of CVs on the sky could be converted to a galactic 
spatial distribution. Previous space density estimates of CVs have been made from the samples 
which include CVs with distances according to various methods. The problems of those early 
estimations are mostly due to inadequate number of CVs with reliable distances. Unfortunately, there 
is no single technique, no ``magic bullet'', for distance finding to CVs \citep{Patterson1998}. 
However, a single method, which is able to provide reliable distances of all CVs, would be crucial 
in computing the true space distribution and the space density of these objects. Such a method was recently 
suggested by \cite{Aketal2007}. The suggested method uses period-luminosity-colours (PCLs) relation 
of CVs calibrated by {\em 2MASS} photometric data and trigonometric parallaxes for a sufficient number of CVs.

Nevertheless, the interstellar reddening is one of the important dilemmas in a distance determination.
The reddening corrections have been made so far by using short-wavelength observations, column 
densities, reddening maps, in general. However, the reddening coefficients in optical wavelengths are 
higher than in infrared wavelengths. The distance estimation technique proposed by \cite{Aketal2007} is 
less affected by interstellar reddening since it is based on infrared colours from {\em 2MASS} observations.

The aim of this paper is to derive the spatial distribution, galactic model parameters and 
luminosity function of CVs in the solar neighbourhood using the distances estimated by 
the new method proposed by \cite{Aketal2007}, which could be applicable to all CVs with reliable 
periods and colours. Consequently, a data sample containing 459 CVs has been collected 
and analysed. Description of data, reddening problem, distances corrected for reddening, 
completeness of sample and galactic model parameters are discussed in section 2. After 
the analysis regarding to space distributions and luminosity functions in Section 3, conclusions 
are summarized in Section 4.

\section{The Data}

Table 1 contains {\em 2MASS} photometric data, colour excesses and orbital periods of CVs which 
were collected for this study. In addition to the columns of name, type, orbital period, galactic 
coordinates ($l, b$), colour excess $E(B-V)$, apperent brightness ($J$) and infrared colours $(J-H)$ 
and $(H-K_{s})$, which are the basic observatinal data, the last two colums (absolute magnitude, 
distance) are added as predictions from the PLCs relation of  \cite{Aketal2007}. 

A preliminary list of cataclysmic variables with orbital periods is collected from 
Downes et al.'s \footnote{http://archive.stsci.edu/prepds/cvcat/} (\citeyear{Downesetal2001}) and
Ritter and Kolb's\footnote{http://www.mpa-garching.mpg.de/RKcat/} (\citeyear{RitterandKolb2003}) 
catalogues. Superhump periods of SU UMa type dwarf novae, for which the orbital periods are not 
known, are accepted as orbital periods. The bias introduced by using the superhump periods should 
be negligible since the superhump periods are only a few per cent longer than the orbital 
periods \citep{Patterson2001}. 
  
The $J$, $H$ and $K_{s}$ magnitudes of the sample stars were taken from the Point-Source Catalogue 
and Atlas \citep{Cutrietal2003,Skrutskieetal2006} which is based on 
the {\em 2MASS} (Two Micron All Sky Survey) observations. The {\em 2MASS} photometric system comprises 
Johnson's $J$ (1.25 $\mu$m) and $H$ (1.65 $\mu$m) bands with the addition of $K_{s}$ (2.17 $\mu$m) 
band, which is bluer than Johnson's $K$-band.  The limiting apparent magnitudes are 15.8, 15.1 and 
14.3, for the three bands respectively \citep{Skrutskieetal2006}. The number of CVs having an orbital 
period and {\em 2MASS} photometric data is 541. However, 82 of them were removed because 
they are beyond the limits of applicability of the PLCs relation, which will be discussed later. 
Consequently, the number of CVs selected for this study was reduced to 459 and all are recorded in Table 1.


\begin{table*}
\begin{center}
\tiny{
\caption{The data sample. Types and orbital periods ($P$, in days) were taken 
from Ritter and Kolb (2003). CV denotes CVs with unknown types, DN dwarf 
novae, NL nova-like stars, N novae and NR recurrent novae. $l$ and $b$ are 
galactic longitude and latitude in degrees, $J$, $(J-H)$ and $(H-K_{s})$ 
{\em 2MASS} observations, $E(B-V)$ colour excess, $M_{J}$ the absolute 
magnitude in $J$-band and $d$ distance in pc.}
\begin{tabular}{lcccccccccc}
\hline
 GCVS-Name &  Type  &   $P$   &  $l$  &  $b$ & $E(B-V)$ &  $J$   & $(J-H)$ & $(H-K_{s})$ &  $M(J)$ &    $d$ \\
\hline
    AM CVn &     NL & 0.01191 &  140  &   79 &   0.013  & 14.482 &  0.039  & -0.203 &  8.69 &  143 \\
    HP Lib &     NL & 0.01276 &  352  &   33 &   0.044  & 13.797 & -0.072  &  0.006 &  9.70 &   65 \\
    CR Boo &     DN & 0.01703 &  341  &   66 &   0.022  & 14.548 &  0.012  & -0.295 &  7.03 &  316 \\
  V803 Cen &     DN & 0.01866 &  309  &   21 &   0.037  & 13.806 & -0.029  & -0.023 &  8.67 &  105 \\
J0926+3624 &     NL & 0.01966 &  188  &   46 &   0.006  & 15.352 &  0.460  &  0.140 & 11.08 &   71 \\
J1427-0123 &     NL & 0.02539 &  346  &   53 &   0.011  & 13.975 &  0.560  &  0.268 & 11.64 &   29 \\
    GP Com &     NL & 0.03234 &  324  &   80 &   0.013  & 15.724 &  0.104  &  0.475 & 11.38 &   73 \\
    EI Psc &     DN & 0.04457 &   90  &  -51 &   0.041  & 14.684 &  0.368  &  0.217 &  9.31 &  117 \\
  V396 Hya &     NL & 0.04521 &  309  &   39 &   0.122  & 16.556 &  0.677  & -0.025 &  8.11 &  465 \\
    GW Lib &     DN & 0.05332 &  341  &   27 &   0.095  & 16.191 &  0.605  &  0.193 &  9.18 &  243 \\
    BW Scl &     DN & 0.05432 &  345  &  -74 &   0.007  & 15.835 &  0.342  &  0.555 & 11.32 &   80 \\
    DI UMa &     DN & 0.05456 &  167  &   43 &   0.015  & 15.532 &  0.270  &  0.136 &  8.00 &  319 \\
  V844 Her &     DN & 0.05464 &   62  &   44 &   0.007  & 16.763 &  0.479  &  0.206 &  9.08 &  343 \\
J0222+4122 &     DN & 0.0548  &  141  &  -18 &   0.015  & 15.401 &  0.626  &  0.284 & 10.01 &  119 \\
    EV UMa &     NL & 0.05534 &  118  &   63 &   0.015  & 16.317 &  0.308  &  0.289 &  9.20 &  264 \\
    GG Leo &     NL & 0.05547 &  232  &   49 &   0.032  & 15.013 &  0.248  &  0.130 &  7.83 &  270 \\
J0154-5947 &     NL & 0.0556  &  289  &  -56 &   0.011  & 15.991 &  0.330  &  0.512 & 10.90 &  104 \\
J1238-0339 &     DN & 0.0559  &  297  &   59 &   0.030  & 16.651 &  0.161  &  0.066 &  7.12 &  798 \\
  V386 Ser &     DN & 0.05592 &   11  &   34 &   0.067  & 16.531 &  0.690  &  0.368 & 10.64 &  147 \\
    HV And &     NL & 0.05599 &  121  &  -19 &   0.072  & 15.372 &  0.226  & -0.016 &  6.58 &  556 \\
    CI Gru &     DN & 0.056   &  359  &  -47 &   0.020  & 16.138 &  0.333  &  0.294 &  9.26 &  236 \\
J0025+1217 &     DN & 0.0562  &  113  &  -50 &   0.063  & 16.663 &  0.729  &  0.357 & 10.65 &  156 \\
    AL Com &     DN & 0.05667 &  283  &   76 &   0.033  & 16.513 &  0.331  &  0.316 &  9.36 &  266 \\
    PU CMa &     DN & 0.05669 &  234  &  -13 &   0.026  & 12.486 &  0.025  &  0.072 &  6.77 &  137 \\
    WZ Sge &     DN & 0.05669 &   58  &   -8 &   0.023  & 14.862 &  0.305  &  0.559 & 11.11 &   56 \\
    SW UMa &     DN & 0.05682 &  165  &   37 &   0.017  & 15.621 &  0.294  &  0.517 & 10.78 &   92 \\
J1839+2604 &     DN & 0.05689 &   55  &   14 &   0.011  & 13.399 &  0.546  &  0.311 &  9.92 &   49 \\
J1959+2242 &     DN & 0.058   &   61  &   -4 &   0.186  & 16.562 &  0.832  &  0.161 &  9.12 &  285 \\
    RZ LMi &     DN & 0.0585  &  191  &   51 &   0.014  & 15.679 & -0.015  &  0.056 &  6.50 &  681 \\
    CC Scl &     DN & 0.0587  &   17  &  -69 &   0.013  & 16.040 &  0.191  &  0.445 &  9.91 &  168 \\
 V1025 Cen &     NL & 0.05876 &  300  &   24 &   0.035  & 15.058 &  0.408  &  0.254 &  9.01 &  160 \\
    EG Cnc &     DN & 0.05877 &  197  &   36 &   0.054  & 16.234 &  0.667  & -0.268 &  5.79 & 1203 \\
T Leo (QZ Vir) & DN & 0.05882 &  263  &   60 &   0.007  & 14.771 &  0.436  &  0.509 & 11.02 &   56 \\
    RW UMi &      N & 0.0591  &  110  &   33 &   0.036  & 16.338 &  0.463  &  0.199 &  8.73 &  327 \\
    FS Aur &     DN & 0.0595  &  181  &    0 &   0.023  & 14.766 &  0.181  &  0.331 &  8.98 &  142 \\
    HT Cam &     NL & 0.05971 &  153  &   31 &   0.030  & 16.000 &  0.024  &  0.388 &  8.97 &  252 \\
    DW Cnc &     NL & 0.05979 &  205  &   22 &   0.012  & 14.654 &  0.318  &  0.303 &  9.14 &  126 \\
 2219+1824 &     DN & 0.0599  &   80  &  -32 &   0.017  & 15.976 &  0.536  &  0.561 & 11.60 &   75 \\
  V587 Lyr &     DN & 0.06    &   70  &    9 &   0.067  & 15.144 &  0.554  &  0.073 &  7.93 &  270 \\
 V1040 Cen &     DN & 0.0603  &  295  &    5 &   0.139  & 16.295 &  0.504  &  0.084 &  7.71 &  492 \\
    KX Aql &     DN & 0.06036 &   50  &   -2 &   0.092  & 15.844 &  1.001  &  0.197 &  9.93 &  147 \\
    FL Cet &     NL & 0.06052 &  155  &  -58 &   0.029  & 16.508 &  0.023  &  0.311 &  8.37 &  419 \\
    AQ Eri &     DN & 0.06094 &  204  &  -25 &   0.068  & 16.425 &  0.764  &  0.057 &  8.32 &  407 \\
    CP Pup &      N & 0.06143 &  253  &   -1 &   0.035  & 14.341 &  0.104  &  0.211 &  7.79 &  201 \\
 V4140 Sgr &     DN & 0.06143 &    1  &  -29 &   0.029  & 16.637 & -0.063  &  0.797 & 11.70 &   96 \\
J1050-1404 &     DN & 0.0615  &  264  &   39 &   0.037  & 16.740 &  0.544  &  0.460 & 10.76 &  155 \\
J0918-2942 &     DN & 0.06165 &  257  &   14 &   0.089  & 16.083 &  0.362  &  0.205 &  8.29 &  349 \\
    CP Tuc &     NL & 0.06183 &  324  &  -54 &   0.013  & 15.695 &  0.674  &  0.270 &  9.74 &  154 \\
 V1141 Aql &     DN & 0.06202 &   41  &   -9 &   0.092  & 16.931 &  1.123  &  0.226 & 10.40 &  195 \\
 V1159 Ori &     DN & 0.06218 &  207  &  -20 &   0.345  & 13.817 &  0.036  &  0.106 &  6.13 &  299 \\
 V2051 Oph &     DN & 0.06243 &  358  &    8 &   0.056  & 14.327 &  0.455  &  0.342 &  9.58 &   87 \\
  V436 Cen &     DN & 0.0625  &  282  &   21 &   0.033  & 14.220 &  0.362  &  0.332 &  9.31 &   94 \\
  V347 Pav &     NL & 0.06255 &  320  &  -26 &   0.037  & 16.201 &  0.679  &  0.537 & 11.64 &   81 \\
    EU UMa &     NL & 0.0626  &  203  &   76 &   0.024  & 16.494 &  0.488  &  0.008 &  7.27 &  694 \\
\hline
\end{tabular}  
}
\end{center}
\end{table*}

\begin{table*}
\contcaption{}
\begin{center}
\tiny{
\begin{tabular}{lcccccccccc}
\hline
 GCVS-Name &  Type  &   $P$   &  $l$  &  $b$ & $E(B-V)$ &  $J$   & $(J-H)$ & $(H-K_{s})$ &  $M(J)$ &    $d$ \\
\hline
    HO Del &     DN & 0.0627  &   58  &  -16 &   0.036  & 13.930 &  0.327  &  0.071 &  7.28 &  211 \\
    VY Aqr &     DN & 0.06309 &   42  &  -35 &   0.057  & 15.278 &  0.423  &  0.267 &  8.91 &  183 \\
J1125-0016 &     DN & 0.0631  &  263  &   56 &   0.010  & 12.395 &  0.694  &  0.134 &  8.74 &   54 \\
    OY Car &     DN & 0.06312 &  290  &  -11 &   0.029  & 14.953 &  0.518  &  0.338 &  9.75 &  108 \\
J1600-4846 &     DN & 0.06338 &  332  &    3 &   0.045  & 13.460 &  0.714  &  0.308 &  9.99 &   48 \\
    MR UMa &     DN & 0.0636  &  163  &   67 &   0.012  & 13.399 &  0.644  &  0.133 &  8.58 &   91 \\
    ER UMa &     DN & 0.06366 &  164  &   48 &   0.004  & 13.606 &  0.152  & -0.041 &  6.03 &  327 \\
Var 79 Per &     DN & 0.0637  &   87  &  -14 &   0.083  & 14.759 &  0.579  &  0.150 &  8.38 &  183 \\
    DO Vul &     DN & 0.064   &   57  &   -4 &   0.071  & 15.602 &  0.313  &  0.315 &  8.92 &  211 \\
    EQ Cet &     NL & 0.06446 &  193  &  -81 &   0.007  & 16.727 &  0.404  &  0.579 & 11.23 &  125 \\
    UV Per &     DN & 0.0649  &  133  &   -4 &   0.098  & 16.468 &  0.742  &  0.295 &  9.79 &  208 \\
    AK Cnc &     DN & 0.0651  &  217  &   32 &   0.035  & 13.772 &  0.017  &  0.039 &  6.15 &  329 \\
    AO Oct &     DN & 0.06535 &  318  &  -34 &   0.067  & 15.503 &  0.481  & -0.131 &  6.02 &  767 \\
J2050-0536 &     NL & 0.06543 &   42  &  -29 &   0.046  & 16.404 &  0.859  &  0.102 &  8.77 &  330 \\
    YY Sex &     NL & 0.0656  &  253  &   45 &   0.044  & 16.476 &  0.928  &  0.157 &  9.35 &  262 \\
  V551 Sgr &     DN & 0.0659  &  357  &   -6 &   0.060  & 14.319 &  0.681  &  0.120 &  8.39 &  150 \\
    IX Dra &     DN & 0.06646 &   97  &   29 &   0.034  & 16.470 &  0.181  &  0.286 &  8.35 &  415 \\
J0209+2832 &     NL & 0.06686 &  143  &  -31 &   0.072  & 16.052 &  0.343  & -0.048 &  6.21 &  903 \\
  V550 Cyg &     DN & 0.0672  &   70  &    0 &   0.093  & 14.592 &  0.426  &  0.182 &  8.06 &  195 \\
    SX LMi &     DN & 0.0672  &  199  &   64 &   0.030  & 15.707 &  0.149  &  0.166 &  7.36 &  462 \\
    SS UMi &     DN & 0.06778 &  106  &   39 &   0.037  & 15.874 &  0.355  & -0.248 &  4.80 & 1614 \\
    BZ UMa &     DN & 0.06799 &  159  &   39 &   0.025  & 14.824 &  0.384  &  0.435 &  9.94 &   94 \\
J0131-0901 &     CV & 0.068   &  153  &  -70 &   0.034  & 15.679 &  0.207  & -0.059 &  5.82 &  926 \\
    KS UMa &     DN & 0.068   &  160  &   52 &   0.006  & 16.087 &  0.447  &  0.573 & 11.16 &   96 \\
    HS Cam &     NL & 0.06821 &  150  &   27 &   0.051  & 16.726 &  0.755  & -0.001 &  7.62 &  648 \\
    EX Hya &     NL & 0.06823 &  303  &   33 &   0.020  & 12.274 &  0.324  &  0.263 &  8.51 &   56 \\
    RZ Sge &     DN & 0.06828 &   56  &   -7 &   0.046  & 15.734 &  0.420  &  0.400 &  9.72 &  157 \\
    TY Psc &     DN & 0.06833 &  131  &  -30 &   0.053  & 13.226 &  0.081  &  0.045 &  6.20 &  249 \\
    IR Gem &     DN & 0.0684  &  187  &   11 &   0.029  & 15.218 &  0.343  &  0.343 &  9.13 &  163 \\
  V699 Oph &     DN & 0.0685  &   10  &   29 &   0.096  & 14.176 &  0.606  &  0.220 &  8.75 &  117 \\
  V393 Pav &     NL & 0.06863 &  340  &  -31 &   0.038  & 16.389 &  0.713  &  0.186 &  8.91 &  309 \\
J2303+0106 &     DN & 0.069   &   76  &  -52 &   0.034  & 16.059 &  0.408  &  0.469 & 10.20 &  147 \\
 V1504 Cyg &     DN & 0.06957 &   76  &   11 &   0.035  & 16.110 &  0.198  &  0.510 &  9.93 &  169 \\
    CY UMa &     DN & 0.06957 &  159  &   59 &   0.009  & 16.012 &  0.512  &  0.469 & 10.50 &  126 \\
    VV Pup &     NL & 0.06975 &  240  &    9 &   0.018  & 15.553 &  0.640  &  0.369 & 10.07 &  124 \\
J1928-5001 &     NL & 0.07016 &  348  &  -26 &   0.030  & 13.411 &  0.467  &  0.151 &  7.97 &  121 \\
    BB Ari &     DN & 0.0703  &  152  &  -29 &   0.090  & 16.551 &  0.634  &  0.234 &  8.88 &  330 \\
  V834 Cen &     NL & 0.0705  &  317  &   16 &   0.015  & 13.467 &  0.311  &  0.391 &  9.35 &   66 \\
J0803+2516 &     DN & 0.071   &  197  &   26 &   0.021  & 14.721 &  0.444  &  0.193 &  8.21 &  199 \\
    PU Per &     DN & 0.0713  &  147  &  -22 &   0.056  & 15.799 &  0.410  &  0.059 &  7.05 &  550 \\
    GD 552 &     DN & 0.07134 &  110  &    4 &   0.098  & 15.385 &  0.587  &  0.340 &  9.48 &  145 \\
    FO And &     DN & 0.07161 &  128  &  -25 &   0.020  & 15.493 & -0.153  &  0.652 & 10.03 &  123 \\
J0953+1458 &     NL & 0.07205 &  220  &   47 &   0.031  & 16.911 &  0.530  &  0.289 &  9.08 &  364 \\
    AW Sge &     DN & 0.0724  &   56  &   -7 &   0.078  & 15.831 &  0.778  &  0.113 &  8.32 &  308 \\
    DT Oct &     DN & 0.0726  &  310  &  -27 &   0.111  & 14.442 &  0.236  &  0.167 &  7.23 &  265 \\
    EP Dra &     NL & 0.07266 &  100  &   24 &   0.061  & 16.274 &  0.582  &  0.196 &  8.45 &  359 \\
    VZ Pyx &     DN & 0.07332 &  250  &   14 &   0.048  & 14.186 &  0.315  &  0.259 &  8.22 &  153 \\
    CC Cnc &     DN & 0.07352 &  204  &   32 &   0.027  & 16.517 &  0.435  &  0.458 & 10.04 &  195 \\
    HT Cas &     DN & 0.07365 &  125  &   -2 &   0.029  & 14.703 &  0.477  &  0.383 &  9.59 &  104 \\
    AY Lyr &     DN & 0.0737  &   67  &   18 &   0.029  & 16.795 &  0.531  &  0.464 & 10.33 &  194 \\
 V1251 Cyg &     DN & 0.07376 &   94  &   -3 &   0.222  & 15.626 &  0.136  &  0.133 &  6.43 &  630 \\
    IY UMa &     DN & 0.07391 &  150  &   52 &   0.008  & 15.725 &  0.625  &  0.235 &  8.92 &  228 \\
J1556-0009 &     DN & 0.07408 &    9  &   38 &   0.072  & 16.285 &  0.544  &  0.475 & 10.34 &  150 \\
    VW Hyi &     DN & 0.07427 &  285  &  -38 &   0.025  & 12.522 &  0.485  &  0.335 &  9.24 &   45 \\
     Z Cha &     DN & 0.0745  &  289  &  -22 &   0.047  & 13.968 &  0.404  &  0.250 &  8.35 &  130 \\
    QW Ser &     DN & 0.07457 &   13  &   49 &   0.024  & 16.274 &  0.325  &  0.557 & 10.46 &  144 \\
J0242-2802 &     DN & 0.0746  &  222  &  -65 &   0.012  & 16.697 &  0.682  &  0.491 & 10.93 &  142 \\
    LY Hya &     DN & 0.0748  &  313  &   32 &   0.031  & 16.137 &  0.463  &  0.466 & 10.12 &  158 \\
    WX Hyi &     DN & 0.07481 &  289  &  -51 &   0.019  & 13.482 &  0.244  &  0.277 &  8.19 &  114 \\
\hline
\end{tabular}  
}
\end{center}
\end{table*}

\begin{table*}
\contcaption{}
\begin{center}
\tiny{
\begin{tabular}{lcccccccccc}
\hline
 GCVS-Name &  Type  &   $P$   &  $l$  &  $b$ & $E(B-V)$ &  $J$   & $(J-H)$ & $(H-K_{s})$ &  $M(J)$ &    $d$ \\
\hline
    BK Lyn &     NL & 0.07498 &  191  &   45 &   0.015  & 14.482 &  0.020  &  0.099 &  6.29 &  432 \\
    FT Cam &     DN & 0.07499 &  140  &    3 &   0.096  & 15.751 &  0.198  &  0.579 & 10.12 &  128 \\
    CE Gru &     NL & 0.0754  &  357  &  -48 &   0.018  & 16.356 &  0.456  &  0.205 &  8.19 &  427 \\
  V893 Sco &     DN & 0.07596 &  348  &   16 &   0.093  & 13.222 &  0.314  &  0.226 &  7.79 &  117 \\
    WY Tri &     DN & 0.07597 &  145  &  -26 &   0.068  & 15.820 &  0.192  &  0.228 &  7.55 &  439 \\
    RZ Leo &     DN & 0.07604 &  265  &   59 &   0.024  & 16.338 &  0.674  &  0.277 &  9.25 &  259 \\
    QY Per &     DN & 0.0761  &  149  &  -13 &   0.119  & 15.347 & -0.078  &  0.164 &  6.25 &  628 \\
     T Pyx &     NR & 0.07622 &  257  &   -1 &   0.088  & 15.149 &  0.187  &  0.278 &  7.85 &  278 \\
 0417+7445 &     DN & 0.07632 &  136  &   17 &   0.051  & 16.014 &  0.599  &  0.418 & 10.03 &  154 \\
    SU UMa &     DN & 0.07635 &  154  &   33 &   0.027  & 11.777 &  0.046  &  0.061 &  6.01 &  141 \\
  V630 Cyg &     DN & 0.0764  &   88  &   -8 &   0.150  & 14.679 &  0.176  &  0.102 &  6.38 &  429 \\
J1730+6247 &     DN & 0.07653 &   92  &   33 &   0.028  & 15.284 &  0.095  & -0.028 &  5.47 &  907 \\
    HS Vir &     DN & 0.0769  &  324  &   52 &   0.037  & 15.016 &  0.146  &  0.267 &  7.75 &  280 \\
    CD Ind &     NL & 0.07701 &  337  &  -41 &   0.018  & 13.821 &  0.590  &  0.266 &  8.93 &   94 \\
    DH Aql &     DN & 0.07738 &   28  &  -12 &   0.154  & 15.932 &  0.669  &  0.039 &  7.15 &  535 \\
  V503 Cyg &     DN & 0.0777  &   82  &    3 &   0.539  & 16.370 &  1.083  &  0.087 &  7.73 &  430 \\
    TY Vul &     DN & 0.0777  &   69  &  -10 &   0.134  & 15.602 &  0.168  &  0.117 &  6.47 &  636 \\
    PV Per &     DN & 0.0778  &  146  &  -20 &   0.040  & 15.181 &  0.036  &  0.193 &  6.88 &  449 \\
  V660 Her &     DN & 0.07826 &   48  &   25 &   0.097  & 14.386 & -0.009  & -0.053 &  4.81 &  791 \\
    BZ Cir &     DN & 0.0784  &  314  &   -8 &   0.169  & 15.612 &  0.540  & -0.089 &  5.81 &  852 \\
J0219-3045 &     DN & 0.0784  &  229  &  -70 &   0.019  & 16.391 &  0.720  &  0.312 &  9.56 &  231 \\
  V884 Her &     NL & 0.07848 &   44  &   19 &   0.030  & 13.785 &  0.497  &  0.270 &  8.64 &  105 \\
    CU Vel &     DN & 0.0785  &  264  &    3 &   0.152  & 14.492 &  0.503  &  0.146 &  7.48 &  237 \\
    MR Ser &     NL & 0.0788  &   32  &   48 &   0.022  & 14.082 &  0.299  &  0.429 &  9.31 &   89 \\
    BL Hyi &     NL & 0.07892 &  296  &  -49 &   0.009  & 14.318 &  0.653  &  0.411 & 10.12 &   69 \\
J2100+0044 &     DN & 0.079   &   50  &  -26 &   0.087  & 16.100 & -0.311  &  0.504 &  8.13 &  378 \\
    ST LMi &     NL & 0.07909 &  212  &   66 &   0.009  & 13.510 &  0.416  &  0.187 &  7.85 &  135 \\
    RS Car &      N & 0.0795  &  291  &   -1 &   0.015  & 15.865 &  1.261  &  0.460 & 12.03 &   58 \\
    BR Lup &     DN & 0.0795  &  334  &   12 &   0.203  & 15.179 &  0.042  &  0.059 &  5.50 &  794 \\
  NSV 9923 &     DN & 0.0796  &  350  &   -9 &   0.103  & 16.092 &  0.243  &  0.160 &  6.98 &  637 \\
    AN UMa &     NL & 0.07975 &  166  &   62 &   0.009  & 15.601 &  0.126  &  0.407 &  8.70 &  239 \\
    WW Hor &     NL & 0.0802  &  272  &  -58 &   0.017  & 16.003 &  0.472  &  0.645 & 11.33 &   86 \\
J0549-4921 &     DN & 0.08022 &  256  &  -30 &   0.044  & 15.619 &  0.409  &  0.341 &  8.86 &  221 \\
    AR UMa &     NL & 0.0805  &  167  &   65 &   0.011  & 14.148 &  0.547  &  0.335 &  9.23 &   96 \\
 V1974 Cyg &      N & 0.08126 &   89  &    8 &   0.191  & 15.672 &  0.389  &  0.078 &  6.51 &  628 \\
    TU Crt &     DN & 0.08209 &  272  &   35 &   0.037  & 16.225 &  0.708  &  0.305 &  9.33 &  236 \\
    HV Aur &     DN & 0.0823  &  166  &   -4 &   0.132  & 13.665 &  0.210  &  0.140 &  6.60 &  245 \\
    QZ Ser &     DN & 0.08316 &   35  &   47 &   0.036  & 15.652 &  0.578  &  0.391 &  9.59 &  161 \\
 V1007 Her &     NL & 0.08328 &   66  &   33 &   0.023  & 14.032 &  0.302  &  0.117 &  6.88 &  267 \\
    TY PsA &     DN & 0.0841  &   26  &  -63 &   0.015  & 14.290 &  0.421  &  0.286 &  8.43 &  148 \\
    BF Ara &     DN & 0.0845  &  344  &   -8 &   0.169  & 14.788 &  0.178  & -0.013 &  5.24 &  758 \\
    KK Tel &     DN & 0.0845  &  346  &  -36 &   0.020  & 12.350 &  0.361  &  0.110 &  6.95 &  119 \\
  V452 Cas &     DN & 0.0846  &  123  &   -9 &   0.118  & 15.908 &  0.305  &  0.304 &  8.02 &  361 \\
J1629+2635 &     NL & 0.0847  &   45  &   42 &   0.043  & 15.458 &  0.625  &  0.163 &  7.97 &  309 \\
  V364 Peg &     DN & 0.085   &   62  &  -24 &   0.051  & 13.211 &  0.322  &  0.074 &  6.51 &  215 \\
J2234+0041 &     DN & 0.085   &   68  &  -47 &   0.064  & 16.449 &  0.425  &  0.453 &  9.54 &  235 \\
    DV UMa &     DN & 0.08585 &  175  &   49 &   0.011  & 16.894 &  1.026  &  0.072 &  8.38 &  502 \\
  V419 Lyr &     DN & 0.0864  &   61  &    9 &   0.106  & 15.600 &  0.347  &  0.164 &  7.07 &  486 \\
    YZ Cnc &     DN & 0.0868  &  194  &   28 &   0.025  & 13.166 &  0.215  &  0.122 &  6.59 &  205 \\
    HU Aqr &     NL & 0.08682 &   45  &  -33 &   0.063  & 14.177 &  0.284  &  0.259 &  7.69 &  194 \\
    IR Com &     DN & 0.08704 &  278  &   83 &   0.031  & 15.032 &  0.421  &  0.029 &  6.41 &  524 \\
    EU Cnc &     NL & 0.0871  &  216  &   32 &   0.018  & 13.328 &  0.464  &  0.042 &  6.65 &  215 \\
J0738+2855 &     CV & 0.0875  &  191  &   22 &   0.024  & 16.564 &  0.633  &  0.438 &  9.99 &  205 \\
  V344 Lyr &     DN & 0.0876  &   73  &   19 &   0.046  & 15.605 &  0.173  &  0.149 &  6.61 &  619 \\
    UZ For &     NL & 0.08787 &  220  &  -53 &   0.012  & 15.317 &  0.420  &  0.121 &  7.11 &  437 \\
J0924+0801 &     NL & 0.088   &  224  &   38 &   0.041  & 16.252 &  0.419  &  0.397 &  9.08 &  268 \\
    GZ Cnc &     DN & 0.0883  &  222  &   36 &   0.038  & 14.211 &  0.108  &  0.122 &  6.23 &  388 \\
J1300-3052 &     DN & 0.08898 &  305  &   32 &   0.072  & 16.465 &  0.480  &  0.288 &  8.33 &  411 \\
    GX Cas &     DN & 0.089   &  123  &   -5 &   0.107  & 16.226 &  0.688  &  0.193 &  8.09 &  405 \\
\hline
\end{tabular}  
}
\end{center}
\end{table*}

\begin{table*}
\contcaption{}
\begin{center}
\tiny{
\begin{tabular}{lcccccccccc}
\hline
 GCVS-Name &  Type  &   $P$   &  $l$  &  $b$ & $E(B-V)$ &  $J$   & $(J-H)$ & $(H-K_{s})$ &  $M(J)$ &    $d$ \\
\hline
    UV Gem &     DN & 0.089   &  195  &    6 &   0.208  & 16.520 &  0.235  &  0.295 &  7.45 &  598 \\
J1556+3523 &     CV & 0.0892  &   57  &   50 &   0.015  & 12.505 &  0.411  &  0.093 &  6.83 &  136 \\
    MT Dra &     NL & 0.08939 &   85  &   23 &   0.040  & 16.086 &  0.444  &  0.300 &  8.39 &  341 \\
    UW Pic &     NL & 0.09264 &  253  &  -33 &   0.029  & 15.625 &  0.337  &  0.401 &  8.79 &  230 \\
  V381 Vel &     NL & 0.0931  &  274  &   13 &   0.137  & 16.619 &  0.437  &  0.074 &  6.39 & 1052 \\
  V725 Aql &     DN & 0.0944  &   50  &   -9 &   0.043  & 11.532 &  0.329  &  0.037 &  6.00 &  125 \\
  V516 Pup &     NL & 0.0952  &  263  &   -9 &   0.063  & 16.255 &  0.307  &  0.638 & 10.32 &  150 \\
    QS Tel &     NL & 0.09719 &  352  &  -27 &   0.031  & 14.293 &  0.524  &  0.244 &  7.99 &  180 \\
    DD Cir &      N & 0.09746 &  311  &   -8 &   0.062  & 15.244 &  0.291  &  0.578 &  9.78 &  121 \\
J1702+3229 &     DN & 0.10008 &   55  &   36 &   0.021  & 15.632 &  0.582  &  0.236 &  8.03 &  328 \\
  V348 Pup &     NL & 0.10184 &  248  &  -12 &   0.155  & 14.865 &  0.125  &  0.180 &  6.10 &  532 \\
    IM Nor &     NR & 0.1026  &  327  &    2 &   0.067  & 15.457 &  1.833  &  0.226 & 11.05 &   74 \\
    MN Dra &     DN & 0.10424 &   99  &   15 &   0.254  & 16.140 &  0.358  &  0.349 &  7.68 &  444 \\
    AP CrB &     NL & 0.10546 &   44  &   50 &   0.027  & 14.714 &  0.416  &  0.410 &  8.74 &  155 \\
     V Per &      N & 0.10713 &  133  &   -5 &   0.182  & 15.950 &  0.393  &  0.095 &  5.98 &  915 \\
  V795 Her &     NL & 0.10825 &   57  &   34 &   0.025  & 12.854 & -0.019  &  0.076 &  5.08 &  354 \\
J0524+4244 &     NL & 0.10915 &  166  &    4 &   0.450  & 16.289 &  0.556  & -0.076 &  4.51 & 1887 \\
J0813+2813 &     NL & 0.11    &  194  &   29 &   0.031  & 16.272 &  0.150  &  0.121 &  5.80 & 1226 \\
  V349 Pav &     NL & 0.1109  &  331  &  -33 &   0.032  & 16.112 & -0.117  &  0.768 &  9.86 &  175 \\
    QU Vul &      N & 0.11177 &   69  &   -6 &   0.052  & 12.504 &  0.693  &  0.164 &  7.44 &  101 \\
  V405 Vul &     DN & 0.112   &   59  &   -3 &   0.108  & 14.531 &  0.452  &  0.196 &  6.93 &  317 \\
J1803+4012 &     NL & 0.1126  &   67  &   26 &   0.023  & 15.237 &  0.477  &  0.363 &  8.40 &  231 \\
J0752+3628 &     NL & 0.114   &  184  &   27 &   0.056  & 15.224 &  0.597  & -0.365 &  3.24 & 2439 \\
  V592 Cas &     NL & 0.11506 &  119  &   -6 &   0.139  & 12.294 &  0.038  &  0.067 &  4.77 &  302 \\
    WX LMi &     NL & 0.11592 &  183  &   58 &   0.006  & 13.427 &  0.594  &  0.340 &  8.50 &   96 \\
J1801-2722 &     DN & 0.117   &    3  &   -2 &   0.083  & 14.024 &  0.935  &  0.221 &  8.32 &  134 \\
    TU Men &     DN & 0.1172  &  289  &  -33 &   0.064  & 14.747 &  0.584  &  0.315 &  8.13 &  205 \\
 V2214 Oph &      N & 0.11752 &  355  &    6 &   0.164  & 15.592 &  0.663  &  0.319 &  8.14 &  289 \\
  V630 Sgr &      N & 0.118   &  358  &   -7 &   0.101  & 14.637 &  0.657  &  0.123 &  6.81 &  352 \\
  V351 Pup &      N & 0.1182  &  253  &   -1 &   0.086  & 16.526 &  0.547  &  0.298 &  7.84 &  527 \\
  V478 Her &     DN & 0.12    &   46  &   30 &   0.069  & 16.047 &  0.492  &  0.205 &  7.01 &  623 \\
 V1084 Her &     NL & 0.12056 &   56  &   40 &   0.017  & 12.459 &  0.041  &  0.080 &  5.02 &  305 \\
    DM Gem &      N & 0.1228  &  185  &   12 &   0.094  & 14.479 &  0.276  &  0.041 &  5.13 &  712 \\
  V442 Oph &     NL & 0.12433 &    9  &    9 &   0.197  & 13.329 &  0.100  &  0.113 &  4.95 &  437 \\
    LQ Peg &     NL & 0.12475 &   66  &  -29 &   0.095  & 14.383 & -0.066  &  0.116 &  4.76 &  809 \\
 V4633 Sgr &      N & 0.12558 &    5  &   -6 &   0.006  & 10.556 & -0.025  &  0.723 &  9.52 &   16 \\
J0230-6842 &     NL & 0.12625 &  290  &  -46 &   0.027  & 16.807 &  1.115  & -0.131 &  6.12 & 1358 \\
    BX Pup &     DN & 0.127   &  242  &    2 &   0.035  & 14.549 &  0.502  &  0.204 &  6.97 &  323 \\
    AH Men &     NL & 0.12721 &  294  &  -28 &   0.089  & 12.478 &  0.406  &  0.178 &  6.40 &  158 \\
    DN Gem &      N & 0.12785 &  184  &   15 &   0.116  & 15.432 &  0.197  & -0.029 &  4.26 & 1634 \\
    KQ Mon &     NL & 0.128   &  227  &    4 &   0.059  & 12.878 &  0.497  &  0.033 &  5.62 &  276 \\
    AM Her &     NL & 0.12893 &   78  &   26 &   0.017  & 11.703 &  0.508  &  0.193 &  6.90 &   91 \\
 V1033 Cen &     NL & 0.13152 &  295  &   -2 &   0.445  & 13.531 &  0.381  &  0.213 &  5.74 &  302 \\
J1610+0352 &     NL & 0.13232 &   16  &   37 &   0.057  & 14.445 &  0.401  &  0.246 &  6.87 &  320 \\
    MV Lyr &     NL & 0.13234 &   75  &   15 &   0.091  & 15.865 &  0.516  &  0.164 &  6.48 &  725 \\
J0837+3830 &     NL & 0.1325  &  184  &   37 &   0.038  & 16.562 &  0.620  &  0.375 &  8.42 &  418 \\
J0809+3814 &     NL & 0.133   &  183  &   31 &   0.048  & 15.288 &  0.089  &  0.319 &  6.60 &  537 \\
 0728+6738 &     NL & 0.13362 &  148  &   29 &   0.034  & 15.138 &  0.330  &  0.285 &  7.00 &  419 \\
 V1494 Aql &      N & 0.13461 &   41  &   -5 &   0.192  & 14.848 &  0.479  &  0.239 &  6.68 &  398 \\
    BG CMi &     NL & 0.13475 &  208  &   13 &   0.032  & 14.550 &  0.268  &  0.143 &  5.77 &  564 \\
    SW Sex &     NL & 0.13494 &  246  &   42 &   0.029  & 14.212 &  0.220  &  0.103 &  5.35 &  584 \\
  V393 Hya &     NL & 0.135   &  276  &   27 &   0.066  & 15.591 &  0.188  & -0.088 &  3.78 & 2244 \\
    HL Aqr &     NL & 0.1356  &   66  &   43 &   0.008  & 13.389 &  0.076  &  0.169 &  5.49 &  378 \\
    OR And &     NL & 0.1359  &  106  &  -10 &   0.179  & 14.167 &  0.031  &  0.108 &  4.55 &  780 \\
    XX Tau &      N & 0.136   &  187  &  -12 &   0.268  & 16.652 &  0.808  & -0.015 &  5.47 & 1547 \\
    DW UMa &     NL & 0.13661 &  150  &   50 &   0.007  & 15.914 &  0.359  &  0.346 &  7.53 &  475 \\
    TT Ari &     NL & 0.13755 &  149  &  -43 &   0.051  & 10.998 &  0.090  &  0.030 &  4.38 &  206 \\
 V2289 Cyg &     NL & 0.138   &   83  &   15 &   0.061  & 15.858 & -0.150  &  0.609 &  8.00 &  364 \\
  V603 Aql &      N & 0.1382  &   33  &    0 &   0.631  & 11.700 &  0.188  &  0.161 &  4.32 &  232 \\
\hline
\end{tabular}  
}
\end{center}
\end{table*}

\begin{table*}
\contcaption{}
\begin{center}
\tiny{
\begin{tabular}{lcccccccccc}
\hline
 GCVS-Name &  Type  &   $P$   &  $l$  &  $b$ & $E(B-V)$ &  $J$   & $(J-H)$ & $(H-K_{s})$ &  $M(J)$ &    $d$ \\
\hline
    DY Pup &      N & 0.139   &  246  &    4 &   0.048  & 14.786 &  0.538  &  0.221 &  6.93 &  365 \\
    WX Ari &     NL & 0.13935 &  164  &  -43 &   0.206  & 14.410 &  0.314  &  0.177 &  5.67 &  514 \\
 V1500 Cyg &      N & 0.13961 &   90  &    0 &   0.350  & 16.125 &  0.633  &  0.116 &  5.73 & 1038 \\
    TT Tri &     NL & 0.13964 &  133  &  -32 &   0.060  & 14.608 &  0.219  &  0.118 &  5.31 &  707 \\
 V1315 Aql &     NL & 0.13969 &   46  &    0 &   0.634  & 14.069 &  0.459  &  0.170 &  5.06 &  489 \\
J0636+3535 &     NL & 0.1397  &  179  &   13 &   0.127  & 15.430 &  0.270  &  0.076 &  4.98 & 1167 \\
    BY Cam &     NL & 0.13975 &  152  &   16 &   0.027  & 13.957 &  0.930  &  0.544 & 10.37 &   52 \\
    BO Cet &     NL & 0.1398  &  162  &  -59 &   0.031  & 14.024 &  0.200  &  0.182 &  5.79 &  438 \\
  V909 Sgr &      N & 0.14    &  359  &  -10 &   0.087  & 15.117 &  0.568  & -0.035 &  5.02 & 1010 \\
 V1223 Sgr &     NL & 0.14024 &    5  &  -14 &   0.078  & 12.810 &  0.071  &  0.099 &  4.73 &  399 \\
 V1432 Aql &     NL & 0.14024 &   29  &  -16 &   0.076  & 14.592 &  0.472  &  0.422 &  8.16 &  187 \\
  V388 Peg &     NL & 0.14063 &   67  &  -35 &   0.066  & 16.204 &  0.521  &  0.144 &  6.25 &  952 \\
 1813+6122 &     NL & 0.1408  &   91  &   28 &   0.037  & 14.835 &  0.123  &  0.067 &  4.71 & 1044 \\
  V849 Her &     CV & 0.1409  &   28  &   35 &   0.073  & 14.951 &  0.192  &  0.101 &  5.06 &  922 \\
    MN Hya &     NL & 0.14124 &  255  &   19 &   0.054  & 14.851 &  0.510  &  0.221 &  6.81 &  397 \\
    AQ Men &     NL & 0.14147 &  292  &  -31 &   0.181  & 13.926 &  0.230  &  0.206 &  5.69 &  412 \\
 V2400 Oph &     NL & 0.142   &    0  &    9 &   0.227  & 13.482 &  0.273  &  0.184 &  5.52 &  356 \\
    AH Pic &     NL & 0.142   &  268  &  -30 &   0.043  & 13.899 &  0.178  &  0.089 &  4.98 &  597 \\
  V584 Lyr &     DN & 0.1429  &   61  &    9 &   0.167  & 14.998 &  0.460  & -0.047 &  4.42 & 1220 \\
    LS Cam &     NL & 0.143   &  141  &   22 &   0.162  & 16.124 &  0.059  &  0.527 &  7.62 &  469 \\
    UU Col &     NL & 0.144   &  236  &  -34 &   0.021  & 16.902 &  0.643  &  0.391 &  8.43 &  491 \\
 V1101 Aql &     DN & 0.1442  &   56  &   56 &   0.021  & 14.898 &  0.434  &  0.041 &  5.30 &  824 \\
    LN UMa &     NL & 0.1444  &  144  &   43 &   0.081  & 14.735 &  0.079  &  0.222 &  5.58 &  656 \\
  V751 Cyg &     NL & 0.14446 &   85  &    0 &   0.341  & 13.622 &  0.193  &  0.192 &  5.09 &  443 \\
    VZ Scl &     NL & 0.14462 &   33  &  -74 &   0.020  & 16.093 &  0.619  &  0.301 &  7.70 &  474 \\
    RR Pic &      N & 0.14503 &  272  &  -26 &   0.046  & 12.458 &  0.060  &  0.144 &  5.02 &  301 \\
    CP Lac &      N & 0.14514 &  102  &   -1 &   0.294  & 15.025 &  0.291  & -0.080 &  3.42 & 1855 \\
  V500 Aql &      N & 0.1452  &   48  &   -9 &   0.062  & 13.901 &  0.517  &  0.200 &  6.59 &  283 \\
J2316-0527 &     NL & 0.14545 &   72  &  -59 &   0.037  & 15.218 &  0.420  &  0.162 &  6.10 &  656 \\
    IM Eri &     NL & 0.14562 &  217  &  -41 &   0.022  & 11.502 &  0.201  &  0.094 &  5.06 &  192 \\
J1007-2017 &     NL & 0.1458  &  259  &   28 &   0.047  & 16.304 &  0.674  &  0.411 &  8.57 &  346 \\
    PX And &     NL & 0.14635 &  117  &  -36 &   0.044  & 14.652 &  0.167  &  0.141 &  5.26 &  742 \\
  V533 Her &      N & 0.147   &   69  &   24 &   0.045  & 14.707 &  0.054  &  0.016 &  4.03 & 1339 \\
 0506+7725 &     NL & 0.1477  &  135  &   21 &   0.148  & 16.259 &  0.525  &  0.074 &  5.44 & 1370 \\
 0455+8315 &     NL & 0.14873 &  129  &   24 &   0.065  & 14.430 &  0.147  &  0.200 &  5.56 &  580 \\
 0220+0603 &     NL & 0.1492  &  160  &  -50 &   0.053  & 15.382 &  0.249  &  0.327 &  6.78 &  515 \\
    BB Dor &     NL & 0.14923 &  268  &  -33 &   0.047  & 14.322 &  0.233  &  0.036 &  4.60 &  864 \\
    VZ Sex &     DN & 0.1493  &  232  &   40 &   0.029  & 14.197 &  0.533  &  0.320 &  7.52 &  214 \\
      Tau2 &     NL & 0.1495  &  184  &  -33 &   0.196  & 14.814 &  0.261  &  0.147 &  5.16 &  787 \\
  V425 Cas &     NL & 0.1496  &  107  &   -6 &   0.178  & 14.134 &  0.156  &  0.123 &  4.75 &  701 \\
    AO Psc &     NL & 0.14963 &   69  &  -53 &   0.066  & 13.459 &  0.154  &  0.217 &  5.68 &  349 \\
    LU Cam &     NL & 0.15    &  146  &   20 &   0.113  & 15.116 &  0.402  &  0.006 &  4.66 & 1178 \\
J0749-0549 &     NL & 0.15    &  225  &   10 &   0.140  & 16.278 &  0.433  &  0.012 &  4.72 & 1932 \\
    EF Tuc &     NL & 0.15    &  310  &  -49 &   0.027  & 13.296 &  0.423  &  0.149 &  5.96 &  290 \\
    AB Dra &     DN & 0.152   &  110  &   23 &   0.111  & 13.623 &  0.331  &  0.186 &  5.78 &  354 \\
    BP Lyn &     NL & 0.15281 &  180  &   42 &   0.013  & 13.854 &  0.108  &  0.219 &  5.65 &  436 \\
  V794 Aql &     NL & 0.1533  &   40  &  -21 &   0.097  & 14.188 &  0.217  &  0.234 &  5.84 &  449 \\
  V992 Sco &      N & 0.15358 &  344  &   -2 &   0.259  & 15.014 &  0.277  &  0.477 &  7.44 &  295 \\
    WY Sge &      N & 0.15364 &   53  &   -1 &   0.177  & 15.693 &  0.860  &  0.387 &  8.47 &  259 \\
    BZ Cam &     NL & 0.15369 &  144  &  -24 &   0.053  & 13.363 &  0.208  &  0.121 &  5.08 &  445 \\
    LD 317 &     NL & 0.154   &  110  &  -18 &   0.079  & 12.892 &  0.178  &  0.135 &  5.05 &  359 \\
 2117-5417 &     NL & 0.1545  &  343  &  -43 &   0.038  & 13.276 &  0.130  &  0.110 &  4.81 &  485 \\
    QQ Vul &     NL & 0.15452 &   62  &   -5 &   0.200  & 13.580 &  0.337  &  0.187 &  5.56 &  370 \\
    PY Per &     DN & 0.1548  &  147  &  -20 &   0.050  & 14.801 &  0.279  &  0.328 &  6.78 &  394 \\
    GS Pav &     NL & 0.15527 &  326  &  -32 &   0.042  & 14.932 &  0.480  &  0.267 &  6.86 &  405 \\
    OY Ara &      N & 0.15547 &  334  &   -4 &   0.269  & 15.567 &  0.506  &  0.191 &  5.87 &  781 \\
    BH Lyn &     NL & 0.15588 &  168  &   35 &   0.040  & 14.812 &  0.208  &  0.176 &  5.48 &  724 \\
    SV CMi &     DN & 0.156   &  212  &   11 &   0.033  & 14.420 &  0.471  &  0.161 &  6.06 &  463 \\
 V1493 Aql &      N & 0.156   &   46  &    2 &   0.044  & 13.555 &  1.126  &  0.376 &  9.33 &   69 \\
\hline
\end{tabular}  
}
\end{center}
\end{table*}

\begin{table*}
\contcaption{}
\begin{center}
\tiny{
\begin{tabular}{lcccccccccc}
\hline
 GCVS-Name &  Type  &   $P$   &  $l$  &  $b$ & $E(B-V)$ &  $J$   & $(J-H)$ & $(H-K_{s})$ &  $M(J)$ &    $d$ \\
\hline
 0642+5049 &     NL & 0.1569  &  165  &   20 &   0.089  & 14.910 &  0.236  & -0.043 &  3.81 & 1598 \\
  V382 Vel &      N & 0.1581  &  284  &    6 &   0.013  & 10.991 & -0.084  &  0.707 &  8.66 &   29 \\
    IP Peg &     DN & 0.15821 &   95  &  -40 &   0.021  & 12.601 &  0.626  &  0.254 &  7.14 &  123 \\
    LX Ser &     NL & 0.15843 &   30  &   51 &   0.042  & 13.926 &  0.159  &  0.121 &  4.90 &  629 \\
    VY For &     NL & 0.1586  &  220  &  -54 &   0.015  & 15.589 &  0.589  &  0.130 &  6.14 &  773 \\
    CY Lyr &     DN & 0.1591  &   57  &   11 &   0.125  & 13.633 &  0.257  &  0.150 &  5.17 &  467 \\
  V380 Oph &     NL & 0.16    &   31  &   16 &   0.168  & 14.157 &  0.147  &  0.173 &  4.95 &  648 \\
 V4077 Sgr &      N & 0.16    &    7  &   -8 &   0.138  & 16.052 &  0.463  &  0.570 &  8.77 &  270 \\
J2337+4308 &     NL & 0.1605  &  109  &  -18 &   0.103  & 15.366 &  0.122  &  0.420 &  6.84 &  486 \\
 0229+8016 &     NL & 0.16149 &  127  &   19 &   0.186  & 13.766 &  0.192  &  0.196 &  5.18 &  484 \\
  V367 Peg &     DN & 0.1619  &   84  &  -36 &   0.057  & 16.423 &  0.481  &  0.492 &  8.39 &  395 \\
    CM Del &     NL & 0.162   &   59  &  -11 &   0.102  & 13.432 &  0.191  &  0.109 &  4.70 &  534 \\
    KT Per &     DN & 0.16266 &  130  &  -11 &   0.121  & 13.311 &  0.488  &  0.198 &  6.08 &  265 \\
    KR Aur &     NL & 0.1628  &  184  &    6 &   0.586  & 16.251 &  0.673  & -0.189 &  2.69 & 4059 \\
    AR And &     DN & 0.163   &  134  &  -23 &   0.047  & 14.589 &  0.593  &  0.266 &  7.01 &  322 \\
    CN Ori &     DN & 0.1632  &  211  &  -15 &   0.311  & 13.806 &  0.501  &  0.203 &  5.73 &  363 \\
    UU Aql &     DN & 0.16353 &   32  &  -18 &   0.085  & 14.262 &  0.692  &  0.239 &  6.98 &  277 \\
    UU Aqr &     NL & 0.16358 &   57  &  -45 &   0.132  & 12.971 &  0.342  &  0.146 &  5.28 &  327 \\
     X Leo &     DN & 0.1644  &  224  &   45 &   0.021  & 14.287 &  0.404  &  0.289 &  6.73 &  322 \\
 V1776 Cyg &     NL & 0.16474 &   83  &    5 &   0.912  & 15.272 &  0.406  &  0.001 &  2.65 & 2301 \\
 V1193 Ori &     NL & 0.165   &  202  &  -21 &   0.128  & 13.547 &  0.073  &  0.077 &  4.06 &  748 \\
    AM Cas &     DN & 0.1652  &  130  &    9 &   0.447  & 12.724 &  0.135  &  0.132 &  3.93 &  479 \\
    DO Dra &     NL & 0.16537 &  130  &   45 &   0.009  & 13.213 &  0.590  &  0.296 &  7.27 &  154 \\
    DO Aql &      N & 0.16776 &   32  &  -12 &   0.245  & 16.145 &  0.554  &  0.277 &  6.49 &  773 \\
    VW Vul &     DN & 0.1687  &   71  &  -13 &   0.132  & 13.524 &  0.250  &  0.106 &  4.67 &  558 \\
 0922+1333 &     NL & 0.169   &  218  &   40 &   0.022  & 13.449 &  0.654  &  0.259 &  7.09 &  186 \\
 0139+0559 &     NL & 0.1692  &  145  &  -54 &   0.047  & 14.915 &  0.071  &  0.218 &  5.21 &  857 \\
    LZ Mus &      N & 0.1693  &  297  &   -3 &   0.157  & 14.455 &  0.707  &  0.241 &  6.79 &  320 \\
J0407-0644 &     DN & 0.17017 &  198  &  -39 &   0.086  & 15.204 &  0.496  & -0.003 &  4.58 & 1286 \\
  V405 Aur &     NL & 0.17262 &  159  &   14 &   0.124  & 13.497 &  0.180  &  0.173 &  4.94 &  489 \\
  V849 Oph &      N & 0.17276 &   39  &   13 &   0.101  & 14.683 &  0.617  &  0.182 &  6.19 &  480 \\
    UZ Ser &     DN & 0.173   &   15  &    2 &   0.173  & 14.031 &  0.481  &  0.349 &  6.91 &  248 \\
  V729 Sgr &     DN & 0.17341 &   12  &  -17 &   0.127  & 13.310 &  0.263  &  0.027 &  4.07 &  670 \\
J0704+2625 &     NL & 0.174   &  190  &   14 &   0.072  & 16.434 &  0.047  &  0.278 &  5.47 & 1514 \\
 V1043 Cen &     NL & 0.17459 &  308  &   30 &   0.039  & 12.817 &  0.665  &  0.220 &  6.71 &  164 \\
    LS Peg &     NL & 0.17477 &   71  &  -30 &   0.066  & 11.675 &  0.088  &  0.093 &  4.21 &  303 \\
J2048+0050 &     NL & 0.175   &   48  &  -25 &   0.097  & 15.628 &  0.607  &  0.174 &  6.08 &  779 \\
  V436 Car &     NL & 0.1753  &  266  &  -14 &   0.159  & 14.639 &  0.246  &  0.289 &  5.86 &  535 \\
    GY Cnc &     DN & 0.17544 &  210  &   39 &   0.029  & 13.958 &  0.566  &  0.274 &  6.86 &  259 \\
    WW Cet &     DN & 0.1758  &   90  &  -71 &   0.030  & 11.075 &  0.144  &  0.108 &  4.53 &  201 \\
    CW Mon &     DN & 0.1766  &  211  &   -3 &   0.112  & 13.874 &  0.534  &  0.319 &  6.91 &  236 \\
     U Gem &     DN & 0.17691 &  199  &   23 &   0.019  & 11.651 &  0.584  &  0.239 &  6.65 &   99 \\
  V405 Peg &     DN & 0.17764 &   94  &  -35 &   0.118  & 12.666 &  0.654  &  0.198 &  6.30 &  179 \\
    ES Dra &     DN & 0.179   &   98  &   47 &   0.018  & 15.458 &  0.578  & -0.009 &  4.78 & 1358 \\
J2216+2900 &     DN & 0.1792  &   87  &  -23 &   0.049  & 14.445 &  0.546  &  0.417 &  7.77 &  212 \\
    BD Pav &     DN & 0.1793  &  338  &  -22 &   0.058  & 13.467 &  0.459  &  0.110 &  5.25 &  430 \\
    GI Mon &      N & 0.1802  &  223  &    5 &   0.107  & 15.334 &  0.317  &  0.205 &  5.46 &  902 \\
 V2306 Cyg &     NL & 0.1812  &   69  &    2 &   0.748  & 15.069 &  0.365  & -0.043 &  2.34 & 2585 \\
    FO Per &     DN & 0.18248 &  151  &    0 &   0.177  & 13.465 &  0.210  &  0.083 &  4.11 &  692 \\
    TW Vir &     DN & 0.18267 &  274  &   55 &   0.027  & 13.350 &  0.270  &  0.141 &  5.01 &  460 \\
    SS Aur &     DN & 0.1828  &  166  &   13 &   0.062  & 12.701 &  0.433  &  0.267 &  6.29 &  187 \\
    MQ Dra &     NL & 0.18297 &   86  &   49 &   0.014  & 14.594 &  0.555  &  0.274 &  6.76 &  368 \\
J0649-0737 &     NL & 0.183   &  220  &   -3 &   0.182  & 15.169 &  0.587  &  0.280 &  6.52 &  498 \\
 0943+1404 &     NL & 0.18416 &  221  &   50 &   0.034  & 15.735 &  0.401  &  0.189 &  5.67 & 1014 \\
      Leo7 &     DN & 0.1868  &  227  &   40 &   0.037  & 14.927 &  0.564  &  0.273 &  6.67 &  441 \\
  V697 Sco &      N & 0.187   &  353  &   -5 &   0.155  & 14.037 &  0.629  &  0.227 &  6.25 &  340 \\
 1857+7127 &     DN & 0.18911 &  102  &   25 &   0.068  & 14.739 &  0.419  &  0.226 &  5.86 &  582 \\
    CG Dra &     DN & 0.1893  &   83  &   19 &   0.052  & 14.783 &  0.522  &  0.231 &  6.18 &  514 \\
    AI Tri &     NL & 0.19175 &  141  &  -30 &   0.061  & 14.210 &  0.286  &  0.300 &  6.03 &  421 \\
\hline
\end{tabular}  
}
\end{center}
\end{table*}

\begin{table*}
\contcaption{}
\begin{center}
\tiny{
\begin{tabular}{lcccccccccc}
\hline
 GCVS-Name &  Type  &   $P$   &  $l$  &  $b$ & $E(B-V)$ &  $J$   & $(J-H)$ & $(H-K_{s})$ &  $M(J)$ &    $d$ \\
\hline
    DQ Her &      N & 0.19362 &   73  &   26 &   0.030  & 13.600 &  0.319  &  0.196 &  5.39 &  433 \\
    IX Vel &     NL & 0.19393 &  265  &   -8 &   0.057  &  9.118 &  0.137  &  0.153 &  4.55 &   80 \\
    CT Ser &      N & 0.195   &   24  &   48 &   0.039  & 16.040 &  0.438  &  0.242 &  6.01 & 1000 \\
  V433 Ara &     NL & 0.19573 &  331  &   -9 &   0.086  & 15.798 &  0.497  &  0.455 &  7.62 &  418 \\
    MU Cam &     NL & 0.19664 &  141  &   24 &   0.095  & 14.341 &  0.225  &  0.263 &  5.47 &  573 \\
    UX UMa &     NL & 0.19667 &  107  &   64 &   0.011  & 12.758 &  0.349  &  0.145 &  5.10 &  339 \\
J1023+0038 &     NL & 0.19809 &  243  &   46 &   0.050  & 16.296 &  0.603  & -0.169 &  3.34 & 3823 \\
  V345 Pav &     NL & 0.1981  &  338  &  -29 &   0.065  & 12.151 &  0.389  &  0.090 &  4.66 &  307 \\
  V895 Cen &     NL & 0.19855 &  323  &   21 &   0.062  & 13.975 &  0.560  &  0.268 &  6.42 &  316 \\
    HX Peg &     DN & 0.2008  &   97  &  -47 &   0.085  & 13.224 &  0.228  &  0.070 &  4.02 &  671 \\
    AT Cnc &     DN & 0.2011  &  199  &   31 &   0.038  & 12.482 &  0.154  &  0.092 &  4.09 &  470 \\
    FO Aqr &     NL & 0.20206 &   53  &  -49 &   0.052  & 12.872 &  0.127  &  0.237 &  5.04 &  361 \\
     T Aur &      N & 0.20438 &  177  &   -2 &   0.232  & 14.016 &  0.277  &  0.162 &  4.46 &  741 \\
J0911+0841 &     DN & 0.2054  &  222  &   35 &   0.062  & 15.332 &  0.628  &  0.189 &  5.93 &  740 \\
  V825 Her &     NL & 0.206   &   66  &   34 &   0.029  & 13.879 &  0.279  &  0.070 &  4.21 &  848 \\
  V446 Her &      N & 0.207   &   45  &    5 &   0.453  & 15.387 &  0.586  &  0.108 &  4.35 & 1342 \\
 V3885 Sgr &     NL & 0.20714 &  357  &  -28 &   0.024  &  9.955 &  0.220  &  0.119 &  4.42 &  127 \\
  V617 Sgr &     NL & 0.20717 &  357  &   -7 &   0.121  & 13.786 &  0.267  &  0.247 &  5.27 &  480 \\
    TW Tri &     DN & 0.20758 &  134  &  -30 &   0.038  & 14.565 &  0.575  &  0.268 &  6.40 &  423 \\
 V4745 Sgr &      N & 0.20782 &    1  &  -12 &   0.108  & 15.446 &  -0.003 &  0.314 &  5.08 & 1132 \\
    RX And &     DN & 0.20989 &  126  &  -21 &   0.028  & 12.454 &  0.707  &  0.187 &  6.14 &  181 \\
    EX Dra &     DN & 0.20994 &   98  &   29 &   0.027  & 12.882 &  0.617  &  0.205 &  6.04 &  231 \\
J0808+3131 &     DN & 0.21    &  190  &   29 &   0.042  & 16.178 &  0.515  &  0.118 &  5.10 & 1617 \\
    HZ Pup &      N & 0.213   &  246  &    1 &   0.371  & 15.910 &  0.275  & -0.022 &  2.69 & 3788 \\
    AP Cru &      N & 0.213   &  301  &   -2 &   0.250  & 15.457 &  0.696  &  0.276 &  6.25 &  628 \\
    PW Vul &      N & 0.2137  &   61  &    5 &   0.226  & 16.140 &  0.643  &  0.312 &  6.42 &  801 \\
    HR Del &      N & 0.21417 &   63  &  -14 &   0.106  & 12.323 &  0.048  &  0.056 &  3.24 &  627 \\
    PQ Gem &     NL & 0.21636 &  206  &   20 &   0.025  & 13.494 &  0.293  &  0.205 &  5.13 &  465 \\
    HL CMa &     DN & 0.21679 &  227  &   -8 &   0.168  & 11.637 &  0.188  &  0.216 &  4.62 &  237 \\
    AY Psc &     DN & 0.21732 &  142  &  -54 &   0.050  & 14.517 &  0.500  &  0.015 &  4.20 & 1134 \\
    CZ Ori &     DN & 0.2189  &  195  &    0 &   0.107  & 12.587 &  0.227  &  0.061 &  3.68 &  578 \\
J2243+3055 &     DN & 0.21894 &   93  &  -24 &   0.064  & 12.874 &  0.091  &  0.057 &  3.40 &  766 \\
J0900+4301 &     DN & 0.221   &  178  &   41 &   0.020  & 15.746 &  0.681  &  0.092 &  5.26 & 1241 \\
  V709 Cas &     NL & 0.2225  &  120  &   -3 &   0.112  & 13.136 &  0.360  &  0.266 &  5.50 &  322 \\
    EZ Del &     DN & 0.2234  &   58  &  -12 &   0.182  & 15.083 &  0.298  &  0.076 &  3.77 & 1701 \\
 V4742 Sgr &      N & 0.225   &    5  &   -1 &   0.519  & 14.458 &  0.442  &  0.283 &  4.91 &  656 \\
  V705 Cas &      N & 0.228   &  114  &   -4 &   0.221  & 14.884 &  0.349  &  0.259 &  5.12 &  820 \\
    TV Col &     NL & 0.2286  &  237  &  -30 &   0.024  & 13.197 &  0.369  &  0.136 &  4.69 &  499 \\
    WX Pyx &     NL & 0.2307  &  245  &   10 &   0.077  & 15.313 &  0.361  &  0.119 &  4.40 & 1477 \\
    RW Tri &     NL & 0.23188 &  147  &  -30 &   0.077  & 11.938 &  0.360  &  0.118 &  4.38 &  316 \\
  V347 Pup &     NL & 0.23194 &  256  &  -27 &   0.036  & 13.129 &  0.648  &  0.192 &  5.75 &  294 \\
    VY Scl &     NL & 0.2323  &   20  &  -72 &   0.021  & 12.839 &  0.049  &  0.091 &  3.48 &  737 \\
J0732-1331 &     NL & 0.2335  &  230  &    3 &   0.079  & 12.906 &  0.369  &  0.082 &  4.12 &  555 \\
    DO Leo &     NL & 0.23452 &  228  &   57 &   0.037  & 15.926 &  0.731  & -0.021 &  4.37 & 2016 \\
    TX Col &     NL & 0.2383  &  247  &  -30 &   0.034  & 13.633 &  0.264  &  0.201 &  4.77 &  585 \\
    AH Eri &     DN & 0.2391  &  208  &  -38 &   0.091  & 15.923 &  0.577  &  0.401 &  6.92 &  609 \\
  M 5-V101 &     DN & 0.242   &    4  &   47 &   0.045  & 15.656 &  0.358  &  0.426 &  6.61 &  633 \\
    RW Sex &     NL & 0.24507 &  252  &   39 &   0.033  & 10.321 &  0.176  &  0.075 &  3.54 &  224 \\
 V1039 Cen &      N & 0.247   &  310  &   -2 &   0.207  & 15.886 &  0.080  &  0.732 &  7.74 &  392 \\
    LL Lyr &     DN & 0.24907 &   67  &   19 &   0.068  & 15.424 &  0.528  &  0.248 &  5.62 &  890 \\
    RU LMi &     DN & 0.251   &  192  &   53 &   0.012  & 16.684 &  0.285  &  0.593 &  7.63 &  643 \\
 V1425 Aql &      N & 0.2558  &   33  &   -4 &   0.293  & 14.538 &  0.476  &  0.202 &  4.58 &  871 \\
    AH Her &     DN & 0.25812 &   45  &   38 &   0.032  & 11.806 &  0.332  &  0.099 &  3.99 &  360 \\
    FY Per &     NL & 0.2585  &  155  &    3 &   0.190  & 11.747 &  0.353  &  0.099 &  3.70 &  377 \\
    TZ Per &     DN & 0.26291 &  134  &   -3 &   0.231  & 13.087 &  0.580  &  0.112 &  4.26 &  531 \\
    TW Pic &     NL & 0.265   &  266  &  -33 &   0.056  & 14.776 &  0.433  &  0.227 &  5.09 &  848 \\
    BV Pup &     DN & 0.265   &  240  &    1 &   0.089  & 13.336 &  0.447  &  0.099 &  4.10 &  677 \\
     U Leo &      N & 0.2674  &  226  &   53 &   0.042  & 16.212 &  0.405  &  0.271 &  5.34 & 1469 \\
    EI UMa &     DN & 0.2681  &  171  &   37 &   0.031  & 13.866 &  0.277  &  0.055 &  3.43 & 1206 \\
\hline
\end{tabular}  
}
\end{center}
\end{table*}

\begin{table*}
\contcaption{}
\begin{center}
\tiny{
\begin{tabular}{lcccccccccc}
\hline
 GCVS-Name &  Type  &   $P$   &  $l$  &  $b$ & $E(B-V)$ &  $J$   & $(J-H)$ & $(H-K_{s})$ &  $M(J)$ &    $d$ \\
\hline
    TT Crt &     DN & 0.26842 &  275  &   47 &   0.028  & 13.874 &  0.480  &  0.208 &  5.10 &  563 \\
    CM Phe &     NL & 0.2689  &  314  &  -65 &   0.011  & 13.047 &  0.581  &  0.229 &  5.54 &  315 \\
    LY Uma &     DN & 0.27128 &  157  &   56 &   0.010  & 13.102 &  0.614  &  0.139 &  4.95 &  426 \\
    SS Cyg &     DN & 0.27513 &   91  &   -7 &   0.055  &  8.516 &  0.160  &  0.057 &  3.03 &  122 \\
 2347-3144 &     NL & 0.277   &   11  &  -76 &   0.013  & 16.135 &  0.602  & -0.013 &  3.74 & 2999 \\
    BY Cir &      N & 0.2816  &  315  &   -3 &   0.613  & 15.663 &  0.300  &  0.391 &  4.58 & 1281 \\
  V426 Oph &     DN & 0.2853  &   33  &   12 &   0.051  & 10.997 &  0.502  &  0.166 &  4.64 &  183 \\
J0813+4528 &     DN & 0.289   &  174  &   33 &   0.049  & 15.937 &  0.684  &  0.106 &  4.64 & 1780 \\
     Z Cam &     DN & 0.28984 &  141  &   32 &   0.020  & 11.571 &  0.526  &  0.189 &  4.90 &  214 \\
    EM Cyg &     DN & 0.29091 &   65  &    4 &   0.106  & 11.735 &  0.402  &  0.183 &  4.34 &  289 \\
  V838 Her &      N & 0.29764 &   43  &    7 &   0.406  & 16.129 &  0.666  &  0.104 &  3.73 & 2559 \\
J2133+5107 &     NL & 0.2997  &   94  &   -1 &   0.253  & 13.884 &  0.337  &  0.089 &  3.08 & 1306 \\
 V2274 Cyg &      N & 0.3     &   73  &    2 &   0.526  & 16.120 &  0.441  &  0.494 &  5.74 &  961 \\
    AC Cnc &     NL & 0.30048 &  214  &   31 &   0.036  & 13.078 &  0.383  &  0.102 &  3.76 &  720 \\
    RY Ser &     DN & 0.3009  &   11  &   13 &   0.314  & 13.699 &  0.631  &  0.181 &  4.38 &  642 \\
 V2069 Cyg &     NL & 0.31168 &   87  &   -6 &   0.449  & 14.347 &  0.431  &  0.182 &  3.48 & 1238 \\
 V2275 Cyg &      N & 0.3145  &   89  &    1 &   0.071  & 16.246 &  0.220  &  1.037 & 10.05 &  169 \\
    HQ Mon &     NL & 0.316   &  214  &    5 &   0.117  & 12.645 &  0.165  &  0.110 &  2.96 &  826 \\
  V363 Aur &     NL & 0.32124 &  172  &    2 &   0.344  & 13.295 &  0.332  &  0.160 &  3.22 &  901 \\
  V392 Hya &     DN & 0.32495 &  275  &   28 &   0.051  & 14.810 &  0.661  &  0.097 &  4.22 & 1283 \\
 V1309 Ori &     NL & 0.33261 &  200  &  -21 &   0.129  & 14.223 &  0.561  &  0.150 &  4.12 &  995 \\
    BT Mon &      N & 0.33381 &  214  &   -3 &   0.438  & 14.399 &  0.442  &  0.236 &  3.77 & 1119 \\
    MU Cen &     DN & 0.342   &  296  &   17 &   0.095  & 13.029 &  0.507  &  0.237 &  4.63 &  460 \\
    CH UMa &     DN & 0.34318 &  143  &   43 &   0.061  & 12.711 &  0.505  &  0.166 &  4.17 &  499 \\
  V368 Aql &      N & 0.3452  &   44  &   -4 &   0.485  & 14.531 &  0.383  &  0.114 &  2.52 & 2068 \\
    GY Hya &     DN & 0.34724 &  329  &   31 &   0.096  & 13.583 &  0.542  &  0.113 &  3.76 &  885 \\
J2256-2743 &     DN & 0.3501  &   25  &  -65 &   0.035  & 15.419 &  0.514  &  0.144 &  4.04 & 1861 \\
  HD 45166 &     CV & 0.357   &  203  &   -2 &   0.044  &  9.812 &  0.061  &  0.181 &  3.06 &  220 \\
    QZ Aur &      N & 0.3575  &  174  &   -1 &   0.100  & 15.829 &  0.543  &  0.315 &  5.18 & 1294 \\
    RZ Gru &     NL & 0.36    &  353  &  -60 &   0.014  & 11.938 &  0.263  &  0.120 &  3.18 &  561 \\
    RU Peg &     DN & 0.3746  &   74  &  -35 &   0.054  & 11.069 &  0.441  &  0.164 &  3.78 &  281 \\
    AT Ara &     DN & 0.3755  &  344  &   -6 &   0.117  & 13.500 &  0.561  &  0.422 &  5.86 &  322 \\
    SY Cnc &     DN & 0.38    &  210  &   36 &   0.027  & 11.273 &  0.222  &  0.129 &  2.98 &  451 \\
J0527-6954 &     CV & 0.3926  &  281  &  -33 &   0.075  & 15.588 &  0.733  & -0.079 &  2.59 & 3865 \\
    NY Lup &     NL & 0.411   &  332  &    7 &   0.247  & 13.225 &  0.477  &  0.220 &  3.64 &  747 \\
    AE Aqr &     NL & 0.41166 &   45  &  -24 &   0.043  &  9.459 &  0.542  &  0.144 &  3.69 &  140 \\
 V1062 Tau &     NL & 0.41284 &  178  &  -10 &   0.415  & 14.324 &  0.639  &  0.449 &  5.37 &  522 \\
    WX Cen &     NL & 0.41696 &  305  &   -1 &   0.161  & 11.863 &  0.342  &  0.361 &  4.48 &  281 \\
 0928+5004 &     NL & 0.41838 &  168  &   46 &   0.013  & 15.653 &  0.252  &  0.112 &  2.73 & 3826 \\
     Q Cyg &      N & 0.4202  &   90  &   -7 &   0.566  & 13.541 &  0.292  &  0.144 &  1.84 & 1736 \\
    KO Vel &     NL & 0.422   &  278  &    7 &   0.178  & 16.649 &  0.703  &  0.282 &  4.77 & 2213 \\
    DX And &     DN & 0.4405  &  108  &  -16 &   0.113  & 13.106 &  0.594  &  0.122 &  3.34 &  858 \\
    CAL 87 &     CV & 0.44268 &  282  &  -31 &   0.075  & 16.389 &  0.571  &  0.053 &  2.84 & 4970 \\
    QU Car &     NL & 0.454   &  294  &    1 &   0.137  & 10.972 &  0.124  &  0.141 &  2.13 &  554 \\
    UY Pup &     DN & 0.47927 &  231  &    6 &   0.187  & 14.145 &  0.514  &  0.136 &  2.87 & 1670 \\
J1951+3716 &     NL & 0.492   &   72  &    5 &   0.209  & 13.250 &  0.550  &  0.147 &  2.93 & 1065 \\
     V Sge &     CV & 0.5142  &   62  &   -9 &   0.070  & 10.315 &  0.202  &  0.302 &  3.36 &  240 \\
    BV Cen &     DN & 0.61011 &  309  &    7 &   0.220  & 11.290 &  0.404  &  0.205 &  2.42 &  544 \\
    CI Aql &     NR & 0.61836 &   32  &   -1 &   0.139  & 13.668 &  0.339  &  0.644 &  5.63 &  382 \\
J1730-0559 &     NL & 0.6425  &   18  &  -15 &   0.108  & 14.384 &  0.526  &  0.211 &  2.89 & 1904 \\
  V723 Cas &      N & 0.69327 &  125  &   -9 &   0.091  & 11.882 &  0.196  &  0.755 &  5.90 &  152 \\
\hline
\end{tabular}  
}
\end{center}
\end{table*}

\subsection{Intrinsic colours and colour excesses}

Intrinsic $(J-H)_{0}$ and $(H-K_{s})_{0}$ colours and an orbital period are the basic parameters needed for 
the PLCs relation of \cite{Aketal2007} to provide an absolute magnitude of a CV, from which its distance 
can be computed. However, unlike the observed colours, which are directly available from the photometric 
observations, derivation of the intrinsic colours requires one more step namely using various colour 
excess values, if spectroscopic observations are not available to reveal them independently. For 
estimating the $(J-H)_{0}$ and $(H-K_{s})_{0}$ colours, for this study the colour excess of 
$E(B-V)$ has been used.

Although the $E(B-V)$ values for some CVs are given in \citet{BruchandEngel1994}, we have 
preferred to use their first order predictions from the Schlegel, Finkbeiner $\&$ Davis 
(\citeyear{Schlegeletal1998}) maps by using NASA Extragalactic 
Database\footnote{http://nedwww.ipac.caltech.edu/forms/calculator.html} in order to obtain 
a self-consistent data set. This is because those and other available colour excess values in the 
literature are unreliable since some of them are only assumed values, rather than a result of 
an investigation. \citet{Schlegeletal1998} maps, on the other hand, provide only the $E(B-V)$ values 
according to galactic coordinates, which are modeled for any direction from Sun to the 
edges of our Galaxy as a consequence of extinction by galactic dust. Those rough colour excess values 
according to galactic latitude ($b$) and longitude ($l$) towards the directions of stars are 
shown by $E_{\infty}(B-V)$, which symbolically means up to infinity but actually up to the edge 
of the Galaxy. Therefore, the $E_{\infty}(B-V)$ values have to be reduced according to the actual 
distance of each star. At first, a total interstellar absorption within the Galaxy in the photometric 
$V$ band was computed from an available modeled value of $E_{\infty}(B-V)$ for a given galactic 
latitude ($b$) as

\begin{equation}
A_{\infty}(b)=3.1E_{\infty}(B-V),
\end{equation}

which could be reduced to give interstellar absorption in the $V$ band up to the distance 
$d$ of a CV as following
\citep{BahcallandSoneira1980}
\begin{equation}
A_{d}(b)=A_{\infty}(b)\Biggl[1-exp\Biggl(\frac{-\mid d~sin(b)\mid}{H}\Biggr)\Biggr],
\end{equation}
where $H$ is the scaleheight for the interstellar dust which is adopted to be 100 pc as usual 
\citep[see e.g.][]{mendez98}. However, the distance of the CV is needed, but it is unknown at 
this point. Nevertheless, a first order approximation of the distance $d$, which is according 
to the PLCs relation of \cite{Aketal2007} relying on the observed $(J-H)$ and $(H-K_{s})$ colours 
in Table 1 instead of true intrinsic colours, could be used in Eq. 2 for a first order 
estimate of the interstellar extinction in the $V$ band. This extinction, then, is converted to 
an $E_{d}(B-V)$ value as following
\begin{equation}
E_{d}(B-V)=A_{d}(b)~/~3.1,
\end{equation}
which gives the first order approximation of the colour excess $E(B-V)$ for the associated CV.
The predicted colour excess $E(B-V)$ values by this method are given in the column 6 of Table 1.
Once, an approximated $E(B-V)$ value is available for a CV, then the interstellar extinctions at 
$J$, $H$ and $K_{s}$ bands for it could be computed as $A_{J}=0.887\times E(B-V)$, 
$A_{H}=0.565\times E(B-V)$ and $A_{K_{s}}=0.382\times E(B-V)$ according to the equations 
of \citet{FiorucciandMunari2003} \citep[see also,][]{BilirGuverAslan2006}. After computing the 
interstellar extinctions $A_{J}$, $A_{H}$ and $A_{K_{s}}$, the de--reddened colours $(J-H)_{0}$ 
and $(H-K_{s})_{0}$ could be obtained from $J_{0}=J- A_{J}$, 
$H_{0}=H- A_{H}$ and $(K_{s})_{0}=K_{s}- A_{K_{s}}$. Those intrinsic colours and the orbital 
periods (column 3) were then used for computing the absolute magnitudes in column 10 of Table 1 
by the PLCs relation of CVs.

\subsection{Accuracy and limitations of the PLCs relation}

The PLCs relation was established as a useful tool for predicting CV 
distances easily from the orbital periods and {\em 2MASS} observations. 
After discovered by \cite{Aketal2007} as an idea, it was calibrated 
and tested by the most reliable trigonometric parallaxes of 27 CVs with $\sigma_{\pi}/\pi< 0.40$. 
Therefore, it is especially preferred for this study over the other CV distance indicators 
because of its consistency, tested reliability and wide ranges of its aplicability. The open 
form of the PLCs relation is 

\begin{equation}
M_{J}=-0.894-5.721~\log P_{orb}+2.598~(J-H)_{0}+7.380~(H-K)_{0},
\end{equation}
which uses intrinsic $(J-H)_{0}$ and $(H-K_{s})_{0}$ colours and orbital periods to predict absolute magnitude 
of a CV. The relation was claimed to provide absolute magnitudes within the accuracy of 
about $\pm$0.22 mag and valid for the  ranges; 
$0.032^{d} < P_{orb} \leq 0.454^{d}$, $-0.08 < (J-H)_{0} \leq 1.54$, 
$-0.03 < (H-K_{s})_{0} \leq 0.56$ and $2.0 < M_{J} < 11.7$,  
covering the present data very well. Only 15 per cent of CVs in the preliminary list is removed 
because of these limits when forming the list of CVs in Table 1.  

Once absolute magnitudes in the $J$ band are available, the distances 
were then calculated by the Pogson's relation as $d(pc)=10^{(J-M_{J}+5-A_{J})/5}$,
where $J$ is observed apparent magnitude and $A_{J}$ is the interstellar absorption in the $J$ band. 
The distances of CVs in our sample, which are computed by the intrinsic colours, are listed in the 
last column of Table 1. 

The intrinsic colours used in the PLCs relation have been computed as a 
first order estimation as described. Refinements have not been attempted since a possible correction 
term and its effect on the predicted distance would be too small and expected to be lost within about 10 per cent 
uncertainty which is equavalent to $\pm$0.22 mag in the magnitude scale.


\begin{figure}
\begin{center}
\includegraphics[scale=0.375, angle=0]{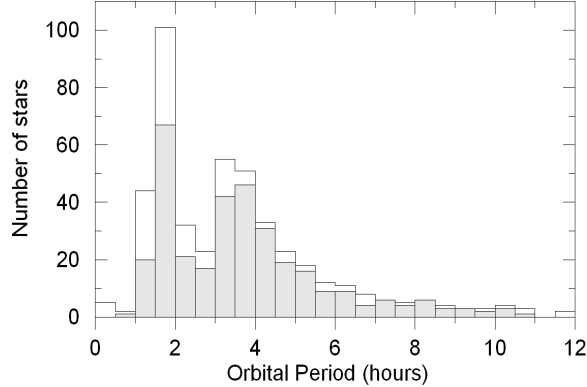}
\caption[] {\small The orbital period distribution of CVs in the sample.}
\end{center}
\end{figure}

\subsection{Completeness of the sample}

Disagreements between the observational and theoretical studies of CV populations could be often 
attributed to the incompleteness of surveys, which are mostly due to unavoidable selection effects of 
observational data  
\citep{DellaValleandLivio1996,Aungwerojwitetal2005,Gansicke2005,PretoriusKniggeKolb2007}.
Indeed, one of the goals of CV surveys is to reduce the selection effects by observing the fainter CVs and 
thereby provide a more accurate picture of the actual CV population since the theory predicts 
the most of CVs should be intrinsically faint \citep{Howelletal1997, Kolb2001}. It is obvious
that the apparent magnitude introduces a strong bias in a CV sample as the detection of faint 
stars is difficult. Moreover, variability and amplitude of the variability are important factors 
which may affect the discovering probability of a CV since rare and low-amplitude erupters are harder 
to discover \citep{Patterson1998,PretoriusKniggeKolb2007}. Thus, low-mass transfer rate systems 
are likely under-represented in the known samples of CVs because the intrinsic brightness and eruption 
frequency decrease with the mass transfer rate. 

Nevertheless, the strongest bias in the present sample may appear to be  originating from the chance of finding 
CVs in the {\em 2MASS} observations. Because this bias mostly depends on the apparent magnitudes, 
it should not be counted as a new kind of bias on the studies which aim to find the space densities 
in the solar neighbourhood. This is because the systems fainter than the limiting magnitude of 
the {\em 2MASS} database are expected to be very distant objects. The limiting magnitudes of 
the {\em 2MASS} survey are 15.8, 15.1 and 14.3 in $J$, $H$ and $K_{s}$ bands, 
respectively \citep{Skrutskieetal2006}. Therefore, we had to remove 114 systems from our primary sample, 
which was including more than 640 systems, as they do not exist in the {\em 2MASS} database 
mostly because they are dimmer than the limiting magnitudes. 
Additionally, we had to remove 46 systems more from the preliminary sample due to the application 
limits of the PLCs relation concerned with the de-reddened colour indices $(J-H)_{0}$ and 
$(H-K_{s})_{0}$, as well (see Section 2.2). 

Another important bias in our sample is introduced by the orbital periods. Although this bias is 
magnitude dependent too, it could be counted as one of the biases that stands out independently because 
it is easier to measure shorter periods. The selection effects due to the orbital periods are more 
pronounced for magnetic systems since these objects are preferred in the follow-up observations 
to measure the orbital period \citep{PretoriusKniggeKolb2007}. It should be noted that the upper 
and lower application limits of the PLCs relation concerned with the orbital period (see Section 2.2) 
do not introduce an additional bias in the CV sample of this study, because almost all CVs with 
known orbital periods are included in the period limits of the PLCs relation. 

Only 24 stars were removed because of the period limits of the PLCs relation. The CVs discovered in 
globular clusters were removed from the sample, as well. After these removing processes, the number of 
systems in the final sample drops to 459. Table 1 contains all of the CVs in the final sample. 
The orbital period distribution of CVs in the final sample is shown in Figure 1. Shaded and white 
areas in Figure 1 show the samples with the limiting apparent magnitudes of $J_{0}$=15.8 and 16.5, 
respectively. Among the all systems (459), 8 are CVs with unknown type, 202 dwarf novae, 57 novae, 
3 recurrent novae and 189 nova-like stars.

Although the CV sample in this study is not free from the selection effects, it is the largest 
sample which has ever been appeared in the literature. Our sample includes $\sim$72 per cent of CVs 
with known orbital periods. Thus, space densities of CVs in the solar neighbourhood derived from 
this sample should be inspiring as well as being useful for constraining the existing population models. 


\begin{figure}
\begin{center}
\includegraphics[scale=0.55, angle=0]{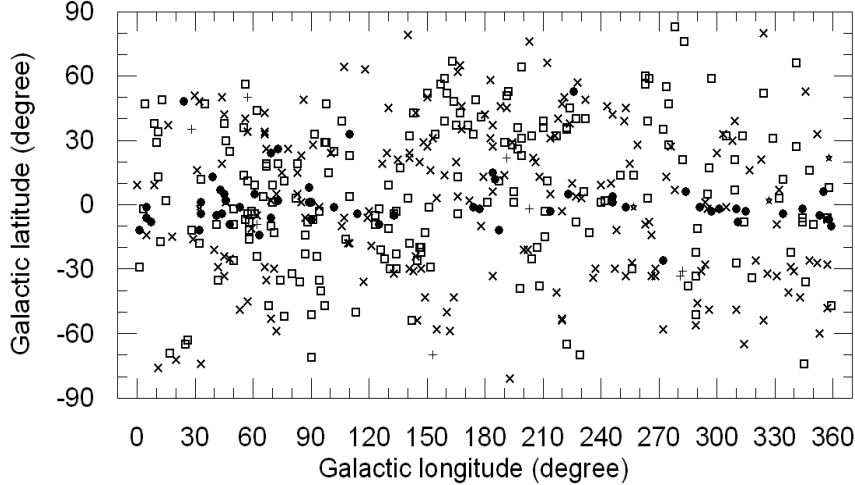}
\caption[] {\small The galactic distribution of CVs in the sample. The symbol ($\square$) denotes dwarf novae, 
($\times$) nova-like stars, ($\bullet$) novae, ($\star$) recurrent novae and ($+$) CVs with unknown type.}
\end{center}
\end{figure}


\begin{figure}
\begin{center}
\includegraphics[scale=0.45, angle=0]{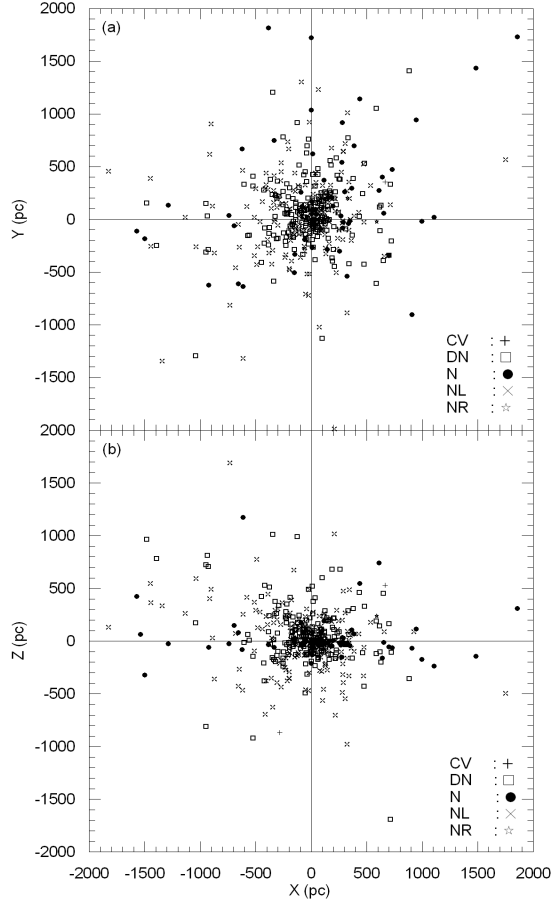}
\caption[] {\small The spatial distribution of systems in the sample with respect to the Sun. 
$X$, $Y$ and $Z$ are heliocentric galactic coordinates directed towards the Galactic Centre, 
galactic rotation and the north Galactic Pole, respectively. CV denotes CVs with 
unknown types, DN dwarf novae, NL nova-like stars, N novae and NR recurrent novae.}
\end{center}
\end{figure}

\subsection{Galactic model parameters}

Exponential functions have usually been used to derive the galactic model parameters 
of the galactic disc. However, a recent study has shown that the observed vertical distribution 
in the Galaxy is smoother in the solar neighbourhood and it is well-approximated by a secans 
hiperbolicus function \citep{Biliretal2006}. 

Therefore, in order to derive scaleheights of CV populations in the solar neighbourhood, we 
have preferred to test following two functions
\begin{equation}
n(z)=n_{0}\exp\Biggl(-\frac{\mid z\mid}{H}\Biggr),
\end{equation}
 and 
\begin{equation}
n(z)=n_{0}sech^{2}\Biggl(-\frac{\mid z\mid}{H_{z}}\Biggr),
\end{equation}
for describing the number density variation of CVs by the distance from the galactic disc.
When choosing the best function to describe actual CV distribution in the solar neigbourhood, the 
galactic model parameters ($n_{0}$ and a scaleheight) were determined for both of the functions first 
and then their fits, that is, abilities of describing actual distribution are compared by a minimum 
$\chi^{2}$ test. Consequently, $n_{0}$ is the number density of CVs in the galactic plane, $z$ is 
the distance from the galactic plane and is given by $z=z_{0}+d\sin(b)$, 
where $b$ is the galactic latitude of the CV and $z_{0}$ is the distance of Sun from the galactic 
plane \citep[24 pc,][]{Juric06}. $H$ and $H_{z}$ are the exponential and $sech^{2}$ vertical 
scaleheights, respectively. The relation between the exponential 
scaleheight $H$ and the $sech^{2}$ scaleheight $H_{z}$ is $H=1.08504H_{z}$ \citep{Biliretal2006}.
Because recent surveys show that the radial scalelength of the thin disc stars is longer 
than 2.25 kpc \citep{Juric06}, we have not attempted to estimate the radial scalelength of CVs since 
$95$ per cent of systems in our sample are closer than $\sim$1 kpc in the $X$--$Y$-plane (see Figure 3). 
All error estimates in the analysis were obtained by changing galactic model parameters until an 
increase or decrease by 2$\sigma$ in $\chi^{2}$ was achieved \citep{Pressetal1997}.

\section{Analysis}

\subsection{Spatial distribution}

The galactic coordinates ($l$, $b$) of CVs in the sample are plotted in Figure 2. 
The figure indicates that CVs in general all distributed about the galactic plane 
symmetrically, except for novae which appear concentrated only in the galactic plane. 

In order to inspect the spatial distribution of the present CV sample of all systems, the Sun centered 
rectangular galactic coordinates ($X$ towards Galactic Center, $Y$ galactic rotation, $Z$ north Galactic 
Pole) were calculated. The projected positions on the galactic plane ($X$, $Y$ plane) and on the plane 
perpendicular to it ($X$, $Z$ plane) are displayed in Figure 3. Having 231 systems with $X\leq0$ and 228 
system with $X>0$, 207 systems with $Y\leq0$ and 252 system with $Y>0$, 219 systems with $Z\leq0$ and 240 
system with $Z>0$, we could conclude that there is not a considerable bias introduced by the projected 
positions of these systems. The numbers, median distances and median heliocentric galactic distances of 
CVs are listed in Table 2.


\begin{table}
\begin{center}
\caption{The numbers, median distances and median heliocentric galactic distances of CVs in 
the sample. Distances ($d$) and positions ($X,Y,Z$) are in pc. CV denotes CVs with 
unknown types, ALL all systems in the sample, DN dwarf novae, NL nova-like stars, N novae 
and NR recurrent novae.}
\begin{tabular}{lccccc}
\hline
Type      & Number & $d$  &   $X$  &   $Y$  &   $Z$ \\
\hline
CV        & 8      & 581  &   80   &   19   &  -23  \\
DN        & 202    & 294  &   -3   &   50   &   12  \\
NL        & 189    & 432  &   -43  &   0    &   20  \\
N         & 57     & 628  &   114  &   50   &  -11  \\
NR        & 3      & 278  &   62   &  -40   &  -5   \\
ALL       & 459    & 377  &   0    &   30   &   9   \\
\hline
\end{tabular}
\end{center}
\end{table}


\begin{figure}
\begin{center}
\includegraphics[scale=0.35, angle=0]{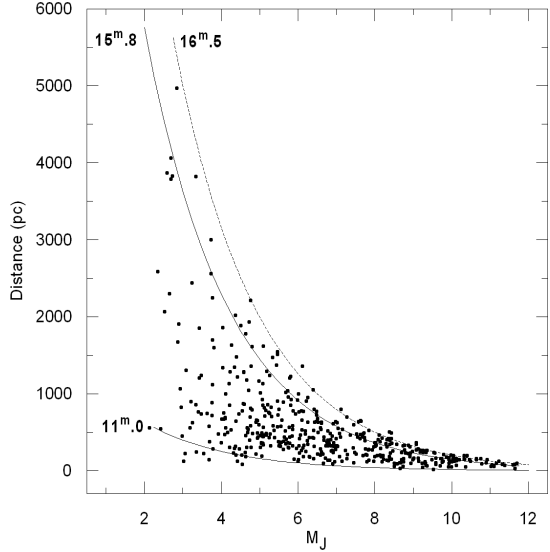}
\caption[] {\small Distances to 459 CVs plotted versus $M_{J}$.
The distance limits for bright ($J_{0}=$11.0) and faint ($J_{0}=$15.8 and 16.5) 
apparent magnitudes are shown by lines.}
\end{center}
\end{figure}

\subsection{Galactic model parameters and space density of CVs}

The completeness limits of the sample according to the apparent magnitudes should be estimated before 
deriving the space density of CVs. Therefore, Figure 4 was drawn to see the CVs distances versus the 
absolute magnitudes $M_{J}$ and then for estimating the completeness limits for bright and faint 
limiting apparent magnitudes in the $J$ band.  

The bright apparent magnitude limit of our sample is consistent with $J_{0}$=11.0. In order to take 
into account the sample's incompleteness towards fainter apparent magnitudes, we assumed two different 
magnitudes, i.e. $J_{0}$=15.8 and 16.5. The limiting magnitude of 15.8 corresponds to the completeness limit 
of $J$ band apparent magnitudes for {\em 2MASS} observations \citep{Skrutskieetal2006}, while the 
limiting magnitude of 16.5 represents almost the whole sample. With the limiting apparent magnitude 
of $J_{0}$=15.8, the number of systems drops to 354. That is, the size of the sample reduces to 
77 per cent of the original size. Consequently, the space densities and galactic model parameters 
of CVs should be examined for both of the limiting apparent magnitudes in order to see how complete the 
present sample is.


\begin{table}
\center
{\scriptsize
\caption{{\scriptsize The galactic model parameters and their errors for CVs. The parameters 
are given for two different apparent magnitude limits. ALL denotes all CVs in the sample, 
DN dwarf novae, NL nova-like stars including polars, AM polars (magnetic systems) and NM 
non-magnetic CVs (all CVs except polars). $n_{0}$ is the number of stars for $z$=0 pc and $H$ 
is the vertical exponential scaleheight. For $sech^{2}$ function, $H$ is calculated from 
$H=1.08504H_{z}$, where $H_{z}$ is the vertical $sech^{2}$ scaleheight.}}
\begin{tabular}{lccccc}
\hline
           &                 & \multicolumn{2}{c}{$J_{0}\leq15.8$}   & \multicolumn{2}{c}{$J_{0}\leq16.5$}\\
Parameter  &  Subgroup       &  $exp$ & $sech^{2}$ & $exp$ & $sech^{2}$                       \\
\hline
$n_{0}$    &  DN             &    111$\pm13$ &  83$\pm8$   &  131$\pm13$  &  100$\pm8$  \\    
$H$(pc)    &                 &    128$\pm20$ &  85$\pm12$  &  150$\pm18$  &  99$\pm12$  \\    
$\chi^{2}_{min}$&            &    1.98       &  4.34       &  3.20        &  2.96       \\
\hline
$n_{0}$    &  NL             &    63$\pm7$   &  49$\pm6$   &  71$\pm7$    &  57$\pm5$   \\    
$H$(pc)    &                 &    250$\pm35$ &  160$\pm5$  &  278$\pm30$  &  177$\pm5$ \\   
$\chi^{2}_{min}$&            &    4.53       &  1.35       &  7.23        &  1.11       \\
\hline 
$n_{0}$    &  AM             &    21$\pm6$   &  17$\pm4$   &  29$\pm5$    &  23$\pm4$   \\    
$H$(pc)    &                 &    185$\pm60$ &  119$\pm35$ &  225$\pm42$  &  142$\pm26$ \\   
$\chi^{2}_{min}$&            &    5.17       &  3.29       &  4.64        &  2.21       \\
\hline  
$n_{0}$    &  NM             &    198$\pm14$ & 151$\pm14$  &  225$\pm15$  &  173$\pm12$ \\    
$H$(pc)    &                 &    154$\pm15$ &  99$\pm11$  &  171$\pm15$  &  110$\pm10$ \\    
$\chi^{2}_{min}$&            &    3.20       &  12.30      &  2.20        &  8.67       \\
\hline 
$n_{0}$    &  ALL            &   218$\pm14$  &  167$\pm13$ &  252$\pm14$  &  195$\pm12$ \\   
$H$(pc)    &                 &   158$\pm14$  &  102$\pm36$ &  179$\pm13$  &  115$\pm9$  \\   
$\chi^{2}_{min}$&            &   2.07        &  15.60      &  1.93        &  6.79       \\
\hline
\end{tabular}
}
\end{table}

The first step of doing this is to prepare $z$-histograms, where $z$ is the distance of CVs 
from the galactic plane computed by equation $z=z_{0}+d\sin|b|$ and binned for per 100 pc 
intervals. The histograms and the best fits of the exponential and $sech^{2}$ functions 
are shown in Figures 5 and 6. The galactic model parameters obtained 
from the minimum $\chi^{2}$ analysis are given in Table 3 for two different faint apparent magnitude 
limits, i.e. 15.8 and 16.5. The numbers and scaleheights of polars (magnetic systems-AM Her stars) 
are estimated separately since the evolution of these systems could be different from the evolution 
of non-magnetic CVs \citep{Wu1993,WebbinkWickramasinghe2002}. 

The number density as a function of $z$ and the scaleheight estimated from the histogram according 
to the magnitude limit 15.8 are smaller than those from the histogram using the limiting magnitude 
of 16.5. This must be due to the fact that faint systems are more distant than brighter ones. 
Figures 5 and 6 show that the exponential function well represents the $z$-histograms of all CVs
for both limiting magnitudes. The vertical scaleheights derived from the exponential function for 
the limiting apparent magnitudes of 15.8 and 16.5 are 158 and 179 pc, respectively. 

However, according to Table 3, the $z$-histograms for the nova-like stars and polars are better 
fitted by $sech^{2}$ functions, while exponential functions give better fits for the other 
two subgroups. Consequently, the better fitting functions require that the scaleheights of 
dwarf novae, nova-like stars, polars and non-magnetic systems are 128, 160, 119 and 154 pc, 
respectively, for the limiting magnitude of 15.8. In the same order, the scaleheights change 
to 150, 177, 142 and 171 for the limiting magnitude of 16.5 (see Table 3).


\begin{figure}
\begin{center}
\includegraphics[scale=0.35, angle=0]{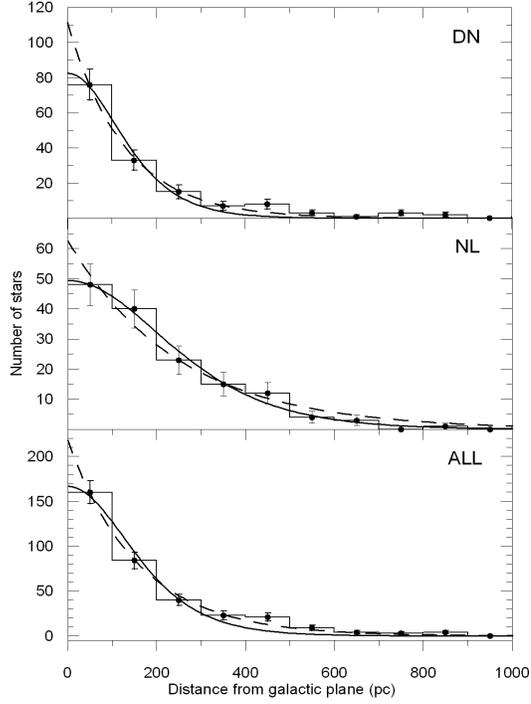}
\caption[] {\small The $z$-histograms for CVs. The systems are selected according to the 
limiting apparent magnitude of $J_{0}$=15.8. Here, ALL denotes all CVs in the sample, 
DN dwarf novae, NL nova-like stars. The dashed line represents the exponential function, 
while the solid line shows the $sech^{2}$ function.}
\end{center}
\end{figure}


\begin{figure}
\begin{center}
\includegraphics[scale=0.35, angle=0]{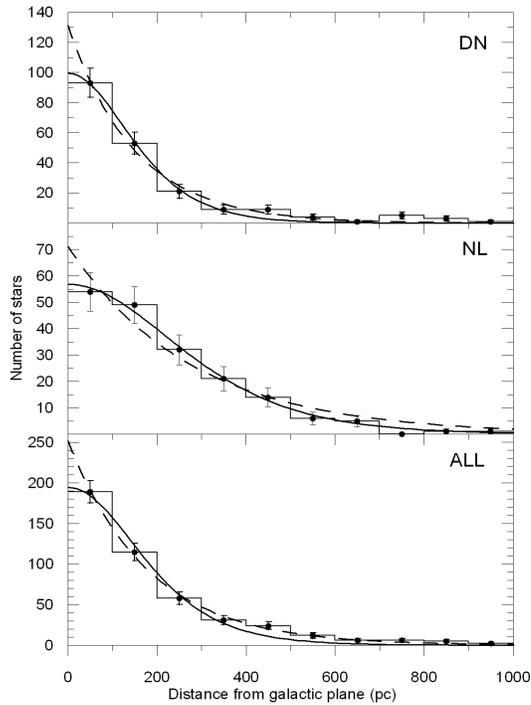}
\caption[] {\small The $z$-histograms for CVs. The systems are selected according to the 
limiting apparent magnitude of $J_{0}$=16.5. The denotes and lines are as in Figure 5.}
\end{center}
\end{figure}


\begin{figure}
\begin{center}
\includegraphics[scale=0.35, angle=0]{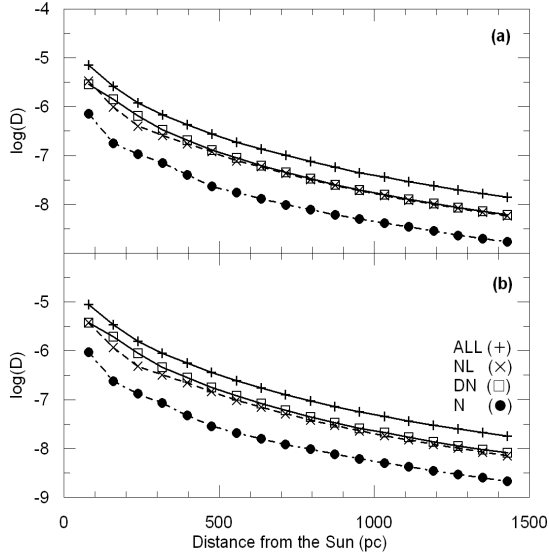}
\caption[] {\small The logarithmic density functions of CVs. The systems are selected according to the 
limiting apparent magnitudes of $J_{0}$=15.8 (a) and 16.5 (b). The denotes are as in Figure 5.}
\end{center}
\end{figure}

Finally, the local space densities are derived by dividing the cumulative number of stars in consecutive 
distances ($d$) from the Sun to the corresponding spherical volumes. The logarithmic density 
functions of dwarf novae, nova-like stars, novae and all stars in the sample are shown in 
Figure 7. The local space densities are summarized in Table 4, which shows that our 
choice of limiting magnitude does not strongly affect the space densities. Moreover, it seems 
that the space density of novae is smaller than those of dwarf novae and nova-like 
stars for a factor of $\sim$1.5 and $\sim$5.8, respectively.


\begin{table}
\begin{center}
\caption{The local space densities of CVs. The results are given for the limiting magnitudes of 
$J_{0}$=15.8 and 16.5.}
\begin{tabular}{lcc}
\hline
                  &        $J_{0}\leq$15.8       &       $J_{0}\leq$16.5          \\
Type              &       $D_{0}$ (pc$^{-3}$)    &      $D_{0}$ (pc$^{-3}$)       \\
\hline
Dwarf novae       &  $6.3(\pm0.6)\times10^{-6}$  &  $7.9(\pm0.9)\times10^{-6}$    \\
Nova-like stars   &  $2.5(\pm0.3)\times10^{-5}$  &  $2.8(\pm0.4)\times10^{-5}$    \\
Novae             &  $4.3(\pm1.2)\times10^{-6}$  &  $5.5(\pm1.4)\times10^{-6}$    \\
All systems       &  $2.9(\pm0.1)\times10^{-5}$  &  $3.2(\pm0.1)\times10^{-5}$    \\
\hline
\end{tabular}
\end{center}
\end{table}

\subsection{Luminosity function of CVs}

The luminosity function is defined as the space density of stars in a certain absolute 
magnitude interval. The logarithmic luminosity functions of CVs according to the limiting 
apparent magnitude of $J_{0}$=15.8 are listed in Table 5 and plotted in Figure 8. 

In Table 5, $d_{1}$ and $d_{2}$ correspond to the distances calculated 
for the lower and upper limiting apparent magnitudes of $J_{0}$=11.0 and 15.8, respectively, 
for an absolute magnitude interval. $\Delta V$ is the shell volume between 
the $d_{1}$ and $d_{2}$ distances and $\phi=\log N/\Delta V$ is the luminosity function.
The luminosity functions are estimated for three types of 
CVs; dwarf novae, nova-like stars, novae and for the combination of all. 
Figure 8 indicates that the luminosity function of dwarf novae and nova-like stars 
are similar, while novae have a rather smaller luminosity function. Note that all 
of the luminosity functions increase towards fainter magnitudes. 


\begin{table}
\center
{\scriptsize
\caption{The logarithmic luminosity functions of CVs in our sample 
with the limiting apparent magnitude of $J_{0}$=15.8. $N$ is the number of stars in 
the $M_{J1}$-$M_{J2}$ absolute magnitude interval and $\phi$ the logarithmic luminosity function.
Distance is in pc, volume pc$^{3}$.}
\begin{tabular}{ccccccccccc}
\hline
&&& \multicolumn{2}{c}{Nova-like stars} & \multicolumn{2}{c}{Dwarf novae} & \multicolumn{2}{c}{Novae} & \multicolumn{2}{c}{All systems}\\
$M_{J1}-M_{J2}$	& $d_{1}-d_{2}$ & $\Delta V$ & 	N   & $\phi$   &  N  & $\phi$& N &  $\phi$  & N & $\phi$\\
\hline
(2.5,3.5]   & 398--3631 & 2.00(11)   & 	12  & -10.22$\pm$0.11  &  5  & -10.60$\pm$0.16 &  4 & -10.70$\pm$0.18  & 22 &  -9.96$\pm$0.08 \\
(3.5,4.5]   & 251--2291	& 5.03(10)   & 	13  &  -9.59$\pm$0.11  & 19  &  -9.42$\pm$0.09 &  7 &  -9.86$\pm$0.14  & 39 &  -9.11$\pm$0.06 \\
(4.5,5.5]   & 158--1445	& 1.26(10)   & 	43  &  -8.47$\pm$0.06  & 17  &  -8.87$\pm$0.09 & 11 &  -9.06$\pm$0.11  & 72 &  -8.24$\pm$0.05 \\
(5.5,6.5]   & 100--912	& 3.17(9)    &	27  &  -8.07$\pm$0.08  & 29  &  -8.04$\pm$0.07 &  7 &  -8.66$\pm$0.14  & 65 &  -7.69$\pm$0.05 \\
(6.5,7.5]   & 63--575	& 7.95(8)    & 	17  &  -7.67$\pm$0.09  & 29  &  -7.44$\pm$0.07 &  8 &  -8.00$\pm$0.13  & 55 &  -7.16$\pm$0.05 \\
(7.5,8.5]   & 40--363	& 2.00(8)    & 	 9  &  -7.35$\pm$0.12  & 19  &  -7.02$\pm$0.09 &  4 &  -7.70$\pm$0.18  & 33 &  -6.78$\pm$0.07 \\
(8.5,9.5]   & 25--229	& 5.02(7)    & 	13  &  -6.59$\pm$0.11  & 14  &  -6.55$\pm$0.10 &  1 &  -7.70$\pm$0.30  & 28 &  -6.25$\pm$0.08 \\
(9.5,10.5]  & 16--145	& 1.28(7)    & 	 5  &  -6.41$\pm$0.16  & 12  &  -6.03$\pm$0.11 &  1 &  -7.11$\pm$0.30  & 18 &  -5.85$\pm$0.09 \\
(10.5,11.5] & 10--91	& 3.15(6)    & 	 2  &  -6.20$\pm$0.23  &  3  &  -6.02$\pm$0.20 &  0 &   	       &  6 &  -5.72$\pm$0.15 \\
\hline
\end{tabular}
}
\end{table}

In order to compare the luminosity functions of DA white dwarfs found from the Anglo 
Australian Telescope Survey \citep[AAT,][]{Boyle1989} and Palomar Green Survey 
\citep[PG,][]{Flemingetal1986} with the luminosity function of CVs in this study, we 
first transformed the $M_{J}$ absolute magnitudes of CVs to Johnson $M_{V}$ absolute 
magnitudes by using Padova Isochrones \citep{Girardietal2002,Bonattaetal2004}. 
In the selection of the isochrones, we have considered the spatial distribution of CVs 
as located in the old thin disc of the Galaxy, in general. Thus, we have obtained an analytical relation 
between $M_{J}$ and $M_{V}$ by assuming a mass fraction of metals of $Z=0.008$, a logarithmic surface 
gravity of $\log g>4$ and a mean age of $t=5.01$ Gyr. The $\chi^{2}_{min}$ analysis show that the best 
fits between the luminosity functions of CVs and white dwarfs are obtained by dividing the luminosity 
functions of white dwarfs to 200 and 400 for the PG and AAT surveys, respectively. Comparison of 
luminosity functions are demonstrated in Figure 9. We conclude that the present luminosity function 
of CVs in this study for $M_{V}>10^{m}$ is reasonably compatible with the luminosity function 
of DA white dwarfs obtained from AAT survey.


\begin{figure}
\begin{center}
\includegraphics[scale=0.45, angle=0]{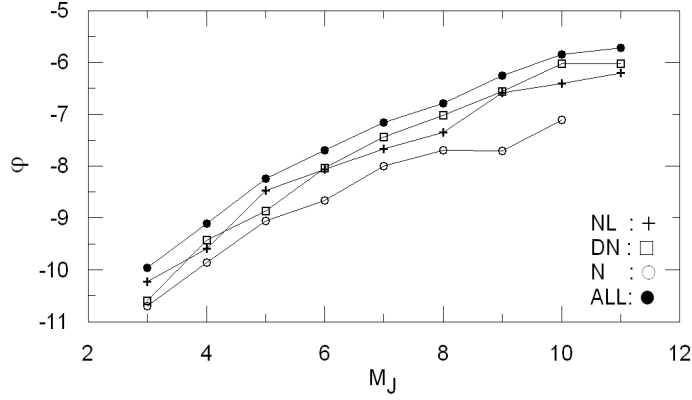}
\caption[] {\small The logarithmic luminosity functions of CVs in our sample 
with the limiting apparent magnitude of $J_{0}$=15.8. The denotes are as in Figure 5.}
\end{center}
\end{figure}


\begin{figure}
\begin{center}
\includegraphics[scale=0.45, angle=0]{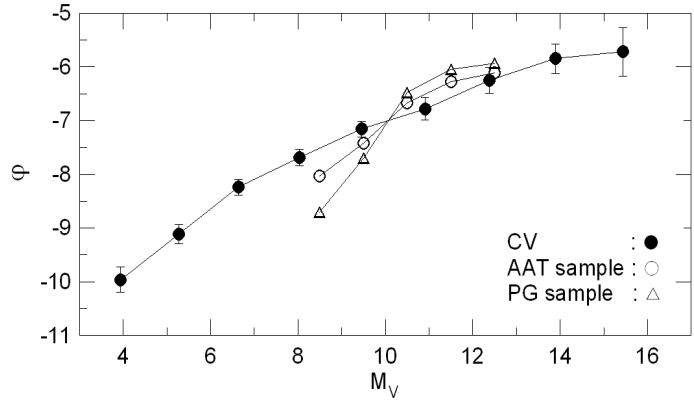}
\caption[] {\small Comparison of the luminosity functions of DA white dwarfs derived from 
the AAT \citep{Boyle1989} and PG surveys \citep{Flemingetal1986}
with the luminosity function of CVs in this study.}
\end{center}
\end{figure}

\section{Conclusions}

The spatial distribution, galactic model parameters and luminosity function of CVs for the 
solar neighbourhood have been derived. In computing the CV distances, a new method suggested 
by \cite{Aketal2007}, which uses orbital periods and intrinsic {\em 2MASS} infrared colours, 
was adopted. This new method advantageously allowed us to collect a fairly homogeneous sample 
of CVs regarding to the distances.

\subsection{Spatial distribution of CVs}

According to Figure 3 and Table 2, most CVs appear to be closer than 1 kpc to the Sun, while 
they are located within the galactic disc with $z$ distances not higher than 0.5 kpc from 
the galactic plane, in general. Novae are strongly concentrated towards the galactic plane. 
Consequently, it might be concluded that because novae share the same galactic regions with the 
young-thin disc population, they could as well  be members of the young population. Such a result 
would be in agreement with that of \citet{Duerbeck1984} who suggested that most novae come from 
fairly young stellar progenitors \citep[see also,][]{Hatanoetal1997,DellaValleandLivio1998}.
As for the spatial distribution of other CV types, the homogeneous spatial distribution of dwarf 
novae and nova-like stars implies that these systems are formed both in young and old populations in the 
galactic disc (see Figure 3). According to \citet{vanParadijsetal1996}, the velocity distribution of 
CVs indicates that they are an old disc population, with a mix of ages up to 10 Gyr. 
Clarification of this point, however, requires further work on the space velocity dispersions 
of CVs and their subgroups. Our forthcoming work will be dedicated to studying the space velocity 
fields and kinematical ages of CVs.

\subsection{Galactic model parameters of CVs}

The $z$-histograms for the nova-like stars and polars are well represented by the $sech^{2}$ functions, 
while the $z$-histograms of dwarf novae are well fitted by an exponential function. In observational 
and theoretical population studies of CVs, usually an exponential distribution function has been assumed. 
However, it should be noted that the observed vertical distribution concerning the stars of all kinds 
in the Galaxy is flat, and better-approximated by a $sech^{2}$ function \citep{Biliretal2006}. 

It has been found in this study that the exponential vertical scaleheights of CVs are 158 and 179 pc for 
the limiting apparent magnitudes of 15.8 and 16.5, respectively. These values are compatible with the 
exponential scaleheights in the range of 100-250 pc and 160-230 pc suggested by \citet{Patterson1984} 
and \citet{vanParadijsetal1996}, respectively. On the other hand, the present data and analysis 
indicate the vertical scaleheights of nova-like stars and polars are within the ranges 160-177 pc 
and 119-142 pc, respectively. These values are very much comparable with the scaleheight of 
$\sim$155 pc determined by \citet{ThomasandBeuermann1998} for the polars. Finally, we have found that 
the exponential vertical scaleheight of dwarf novae is 128 and 150 pc for the two limiting apparent 
magnitudes. The first of these values is in agreement with the scaleheight of 119$\pm$9 pc given 
by \citet{Duerbeck1984}.

\citet{Schwopeetal2002} claimed that 10 to 100 undetected CVs should be hiding in the local vicinity 
(50 pc) of the Sun. According to our analysis, if the $z$ distribution of CVs is modeled by 
an exponential function, there should be about 50 undetected systems in the solar neighbourhood. 
However, if the $z$ distribution of CVs is modeled by a $sech^{2}$ function, as indicated in 
Figures 5 and 6, we would only miss at most 10 CVs in the solar neighbourhood. Consequently, we 
could claim that the discrepancies between the theoretical and observational population studies 
of CVs would almost be removed if the $sech^{2}$ density function is accepted in the theoretical 
population studies.

\subsection{Space density of CVs}

The local space densities calculated according to the limiting apparent magnitude of 15.8 
define a lower limit. However, the local densities do not change considerably when we use the 
limiting apparent magnitude of 16.5, which means that our estimates cannot deviate very much from 
the true space densities.

According to present sample and its analysis the local space density of CVs is  
$2.9(\pm0.1)\times10^{-5}$ pc$^{-3}$. However, former observational population studies of CVs gave predictions 
of space densities from $10^{-6}$ up to $3\times10^{-5}$ pc$^{-3}$ 
\citep{Duerbeck1984,Patterson1998,Hertzetal1990,Schwopeetal2002,Grindlayetal2005}. 
On the other hand, the theoretical population studies of CVs predicted local space densities 
between $10^{-5}$ and $10^{-4}$ pc$^{-3}$ 
\citep{RitterandBurkert1986,deKool1992,Politano1996}. Furthermore, \citet{Sharaetal1986} argued 
that surveys for CVs are very incomplete, and that the space density of CVs could well 
be $10^{-4}$ pc$^{-3}$. However, \citet{Kolb2001} claimed that the space density 
should be smaller than $10^{-4}$ pc$^{-3}$ if the initial mass ratio distribution is more sharply 
peaked to unity. Thus, the intrinsic space density of CVs could well be close to the observational 
value derived here $\sim$$3\times10^{-5}$ pc$^{-3}$. The local space density of CVs predicted from the present 
sample confirms Kolb's (\citeyear{Kolb2001}) theoretical work. 

The local space density of dwarf novae was found to be $6.3(\pm0.6)\times10^{-6}$ pc$^{-3}$ (Table 4). 
This value is in agreement with the space density of $6\times10^{-6}$ pc$^{-3}$ found 
by \citet{Ringwald1993}, who obtained a higher space density than  the observational results of      
\citet{Duerbeck1984} and \citet{Downes1986}. For the nova-like stars in the present CV sample, we 
have derived a space density of $2.6(\pm0.3)\times10^{-5}$ pc$^{-3}$. This value is about 
$\sim$60 times higher than the space density of nova-like stars estimated by \citet{Downes1986}.
On the other hand, space density of novae in the present CV sample 
is $4.3(\pm1.2)\times10^{-6}$ pc$^{-3}$. This could be interpreted as that the space density derived 
in this study is in agreement with  $2.2-4.4\times10^{-6}$ pc$^{-3}$ found by \citet{Patterson1984}, 
which is considerably higher than the space densities of $1.4\times10^{-7}$ 
and $7\times10^{-7}$ pc$^{-3}$ estimated by \citet{Downes1986} and \citet{dellaValleandDuerbeck1993}, respectively.
One of the most important challenges of predicting the space densities is the prediction of 
the true distances. Relying on a homogenized distance prediction, we could claim the space densities 
found in this study are more reliable than in the previous studies.

\subsection{Luminosity function of CVs}

The luminosity function of CVs increases towards fainter absolute magnitudes. Similar trends are 
clear for all types of CVs studied in the present sample. The luminosity function of CVs showing 
such a trend was first examined by \citet{Ringwald1993}. This trend clearly implies an increase 
in the number of short-period systems. This is a result which is roughly consistent with the 
theoretical prediction of \citet{Kolb1993}, who suggested that $\sim$99 per cent of all CVs 
should be intrinsically below the period gap. The theory predicts that most of CVs should 
be intrinsically faint \citep{Howelletal1997}.

We have compared the luminosity function of CVs in the present study with the luminosity functions 
of DA white dwarfs found from the AAT Survey by \citet{Boyle1989} and 
PG Survey by \citet{Flemingetal1986} \citep[see also,][]{Huetal2007}.
Our analysis show that the best fit between the luminosity functions of CVs and DA white dwarfs can be obtained 
by dividing the luminosity function of DA white dwarfs by 400 for the AAT surveys. 
We conclude thereby that one CV is formed for every 400 DA white dwarfs in the solar neighbourhood.
However, it should be noted that it is not clearly known if CV white dwarfs share the narrow mass distribution 
of isolated DA degenerates \citep{Sion1999}.

\section{Acknowledgments}

We thank the anonymous referee for a thorough report and useful comments that helped 
improving an early version of the paper. Part of this work was supported by the Research 
Fund of the University of Istanbul, Project Numbers: BYP-723/24062005 and BYP-1379.
This research has made use of the SIMBAD database, operated at CDS, Strasbourg, France.
This publication makes use of data products from the Two Micron All Sky Survey, which is a joint 
project of the University of Massachusetts and the Infrared Processing and Analysis Center/California 
Institute of Technology, funded by the National Aeronautics and Space Administration and the National 
Science Foundation. This research has made use of the NASA/IPAC Extragalactic Database (NED) which 
is operated by the Jet Propulsion Laboratory, California Institute of Technology, under contract with 
the National Aeronautics and Space Administration. This research has made use of NASA's Astrophysics 
Data System.

\end{document}